\documentclass[aps,prd,superscriptaddress,showpacs,preprint,amsmath,amssymb]{revtex4}
\usepackage{graphicx, bm}
\usepackage{float}
\usepackage[usenames]{color}

\begin{document}

\draft
\title{Feasibility at the LHC, FCC-he and CLIC for sensitivity estimates on anomalous $\tau$-lepton couplings}

\author{ A. Guti\'errez-Rodr\'{\i}guez\footnote{alexgu@fisica.uaz.edu.mx}}
\affiliation{\small Facultad de F\'{\i}sica, Universidad Aut\'onoma de Zacatecas\\
         Apartado Postal C-580, 98060 Zacatecas, M\'exico.\\}

\author{M. K\"{o}ksal\footnote{mkoksal@cumhuriyet.edu.tr}}
\affiliation{\small Deparment of Optical Engineering, Sivas Cumhuriyet University, 58140, Sivas, Turkey.\\}

\author{A. A. Billur\footnote{abillur@cumhuriyet.edu.tr}}
\affiliation{\small Deparment of Physics, Sivas Cumhuriyet University, 58140, Sivas, Turkey.\\}

\author{ M. A. Hern\'andez-Ru\'{\i}z\footnote{mahernan@uaz.edu.mx}}
\affiliation{\small Unidad Acad\'emica de Ciencias Qu\'{\i}micas, Universidad Aut\'onoma de Zacatecas\\
         Apartado Postal C-585, 98060 Zacatecas, M\'exico.\\}

\date{\today}

\begin{abstract}

In this paper, we present detailed studies on the feasibility at $pp$, $e^-p$ and $e^+e^-$ colliders for model-independent
sensitivity estimates on the total cross-section and on the anomalous $\tau^+\tau^-\gamma$ interaction through the tau
pair production channels $pp \to p\tau\bar \tau \gamma p$, $e^-p \to e^- \tau\bar \tau \gamma p$ and $e^+e^-
\to e^+ \tau\bar \tau \gamma e^-$ at the $\gamma^*\gamma^* \to \tau^+\tau^-\gamma$ mode. Measurements of the anomalous
couplings of the $\tau$-lepton $\tilde{a}_\tau$ and $\tilde{d}_\tau$ provide an excellent opportunity to probing extensions
of the Standard Model. We estimate the sensitivity at the $95\%$ Confidence Level, and we consider that the $\tau$-lepton
decays leptonically or semi-leptonically. We found that of the three considered colliders, the future CLIC at high energy and
high luminosity should provide the best sensitivity on the dipole moments of the $\tau$-lepton $\tilde a_\tau= [-0.00128, 0.00105]$
and $ |\tilde{d}_\tau({\rm ecm})|= 6.4394\times 10^{-18}$, which show a potential advantage compared to those from LHC and FCC-he.

\end{abstract}


\vspace{5mm}

\maketitle

\section{Introduction}

The study of the $\tau$-lepton by the ATLAS and CMS Collaborations \cite{ATLAS-tau,ATLAS1-tau} at the Large Hadron Collider (LHC)
has developed significantly in recent years to the point where they have a very active physical program. Furthermore, with the
existence of a $M_H=125.18\pm 0.16\hspace{0.8mm}GeV$ \cite{Data2018} scalar boson \cite{Englert,Higgs,Higgs1,Guralnik} established
by the ATLAS \cite{Atlas} and CMS \cite{CMS} experiments, making it possible to complete the Standard
Model (SM) of particle physics, that is the theory that describes the particles of matter we know, and their interactions. However,
there are fundamental problems in the SM like: dark matter, dark energy, hierarchy problem, neutrino masses, the asymmetry between
matter and antimatter, etc.. These problems demand the construction of new machines that operate at a much higher energy than the LHC,
with cleaner environments and that allow exploring other components of the universe. For these and other reasons, the scientific
community of High Energy Physics has the challenge of discovering that the universe is made in its entirety.

There are several proposals to build new, powerful high-energy and high-luminosity hadron-hadron ($pp$), lepton-hadron ($e^-p$)
and lepton-lepton $(e^+e^-$) colliders in the future at CERN for the post LHC era that will open up new horizons in the field
of fundamental physics.

The future $ep$ colliders, such as the Large Hadron Electron Collider (LHeC) \cite{Klein,Fernandez,Bruening} and the
Future Circular Collider Hadron Electron (FCC-he) \cite{FCChe,Fernandez,Fernandez1,Fernandez2,Huan,Acar,Bruening}, are a
hybrid between the $pp$ and $e^+e^-$ colliders, and they will complement the physical program of the LHC. These colliders have
the peculiarity that can be installed at a much lower cost than that of the $pp$ collider. Furthermore, they provide invaluable
information on the Higgs and top sectors, as well as of others heavy particles as the $\tau$-lepton. The FCC-he study puts great
emphasis in the scenarios of high-intensity and high energy frontier colliders. These colliders, with its high precision and
high-energy, could extend the search of new particles and interactions well beyond the LHC. In addition, in comparison with the
LHC, the FCC-he has the advantage of providing a clean environment with small background contributions from QCD strong interactions.
In the case of the future $e^+e^-$ collider as the Compact Linear Collider (CLIC) \cite{Abramowicz}, although with a much lower
center-of-mass energy than the $pp$ colliders, is ideal for precision measurements due to very low backgrounds.

In this paper we have based our study on three phenomenological analyses for finding physics beyond the Standard Model (BSM) at
present and future colliders to be able to compare the electromagnetic properties of the $\tau$-lepton. We consider $pp$ collisions
at the LHC with 13, 14 TeV and luminosities 10, 30, 50, 100, 200 $fb^{-1}$. Another scenario is the FCC-he with 7.07, 10 TeV and
${\cal L}= 100, 300, 500, 700, 1000\hspace{0.8mm}fb^{-1}$. The CLIC at CERN is another option with 1.5, 3 TeV and luminosities
100, 300, 500, 1000, 1500, 2000, 3000 $fb^{-1}$ have been assumed. With a large amount of data and collisions at the TeV scale,
LHC, FCC-he and CLIC provide excellent opportunities to model-independent sensitivity estimates on the total cross-section of the
production channels $pp \to p\tau\bar \tau \gamma p$, $e^-p \to e^- \tau\bar \tau \gamma p$ and $e^+e^- \to e^+ \tau\bar \tau \gamma e^-$,
as well as of the Magnetic Dipole Moment (MDM) and Electric Dipole Moment (EDM) of the $\tau$-lepton $\tilde a_\tau$ and $\tilde{d}_\tau$.

\begin{table}[!ht]
\caption{Experimental results for the magnetic and electric dipole moment of the $\tau$-lepton.}
\begin{center}
\begin{tabular}{|c| c| c| c|}
\hline
{\bf Collaboration}  &    {\bf Best present experimental bounds on $a_\tau$}            & {\bf C. L.}  &  {\bf Reference}\\
\hline\hline
DELPHI       &   $-0.052 < a_\tau < 0.013$                    & $95 \%$  & \cite{DELPHI} \\
\hline
L3           &   $-0.052 < a_\tau < 0.058$                    & $95 \%$  & \cite{L3} \\
\hline
OPAL         &   $-0.068 < a_\tau < 0.065$                    & $95 \%$ & \cite{OPAL} \\
\hline
\hline
{\bf Collaboration}   &     {\bf Best present experimental bounds on $d_\tau$}          &    {\bf C. L.}  &  {\bf Reference}\\
\hline
\hline
BELLE       &     $-2.2 < Re(d_\tau(10^{-17}e cm)) < 4.5$   &   $95 \%$  & \cite{BELLE} \\
            &     $-2.5 < Im(d_\tau(10^{-17}e cm)) < 0.8$   &   $95 \%$  &               \\
\hline
DELPHI      &     $-0.22 < d_\tau(10^{-16}e cm) < 0.45$     &   $95 \%$  & \cite{DELPHI} \\
\hline
L3          &     $ |Re(d_\tau(10^{-16}e cm))| < 3.1$       &   $95 \%$  & \cite{L3} \\
\hline
OPAL        &     $ |Re(d_\tau(10^{-16}e cm))| < 3.7$       &   $95 \%$  & \cite{OPAL} \\
\hline
ARGUS       &     $ |Re(d_\tau(10^{-16}e cm))| < 4.6$       &   $95 \%$  & \cite{ARGUS} \\
            &     $ |Im(d_\tau(10^{-16}e cm))| < 1.8$       &   $95 \%$  &                  \\
\hline
\hline
\end{tabular}
\end{center}
\end{table}

The theoretical prediction on the MDM of the $\tau$-lepton in the SM is well known with several digits \cite{Passera1}:

\begin{equation}
\mbox{SM}:
a_\tau = 0.00117721(5),
\end{equation}

\noindent while the DELPHI \cite{DELPHI}, L3 \cite{L3}, OPAL \cite{OPAL}, BELLE \cite{BELLE} and ARGUS \cite{ARGUS}
Collaborations report the current experimental bounds on the MDM and the EDM in Table I.

The best experimental results on the MDM and the EDM are reported by the DELPHI and BELLE collaborations using the following processes
$e^+e^- \to e^+ \gamma^*\gamma^* e^- \to e^+e^-\tau^+\tau^-$ and $e^+e^- \to \tau^+\tau^-$, respectively. The EDM of the $\tau$-lepton,
is a very sensitive probe for CP violation induced by new CP phases BSM \cite{Yamanaka1,Yamanaka2,Engel}. It is worth mentioning that
the current Particle Data Group limit was obtained by DELPHI Collaboration \cite{DELPHI} using data from the total cross-section
$e^+e^- \to e^+ \gamma^*\gamma^* e^- \to e^+e^-\tau^+\tau^-$ at LEP2.

The MDM and EDM of the $\tau$-lepton allow a stringent test for new physics and have been deeply investigated by many authors,
see Refs. \cite{Bernreuther0,Iltan1,Dutta,Iltan2,Iltan3,Iltan4,Gutierrez1,Gutierrez2,Gutierrez3,Gutierrez4,Ozguven,Billur,Sampayo,
Passera1,Eidelman1,Koksal3,Arroyo1,Arroyo2,Xin,Pich,Atag1,Lucas,Passera2,Passera3,Bernabeu,Bernreuther,Koksal1,Aranda,Fomin}
for a summary on sensitivities achievable on the anomalous dipole moments of the $\tau$-lepton in different context.

A direct comparison between Eq. (1) and the results given in Table I clearly shows that the experiment is far from determining
the anomaly for the MDM of the $\tau$-lepton in the SM. It is therefore of great interest to investigate and propose mechanisms
model-independent to probe the dipole moments of the $\tau$-lepton with the parameters of the present and future colliders, i.e.
the LHC, FCC-he and CLIC, rendering such an investigation both very interesting and timely.

The outline of the paper is organized as follows: In Section II, we introduce the $\tau$-lepton effective electromagnetic interactions.
In Section III, we show sensitivity estimates on the total cross-section and the $\tau$-lepton MDM and EDM through
$pp \to p\tau\bar \tau \gamma p$ at the LHC, $e^-p \to e^- \tau\bar \tau \gamma p$ at the FCC-he and $e^+e^- \to e^+ \tau\bar \tau \gamma e^-$
at the CLIC. Finally, we present our conclusions in Section IV.

\section{The effective Lagrangian for $\tau$-lepton electromagnetic dipole moments}

We following Refs. \cite{Escribano1,Escribano2,Eidelman1}, in order to analyze in a model-independent manner the total cross-section
and the electromagnetic dipole moments of the $\tau$-lepton through the channels $pp \to p\tau\bar \tau \gamma p$ at the LHC,
$e^-p \to e^- \tau\bar \tau \gamma p$ at the FCC-he and $e^+e^- \to e^+ \tau\bar \tau \gamma e^-$ at the CLIC and using the effective
Lagrangian description. This approach is appropriate for describing possible new physics effects. In this context, all the heavy degrees
of freedom are integrated out leading to obtain the effective interactions with the SM particles spectrum. Furthermore, this is justified
due to the fact that the related observables have not shown any significant deviation from the SM predictions so far. Thus, below we describe
the effective Lagrangian we use with potential deviations from the SM for the anomalous $\tau^+ \tau^-\gamma$ coupling and fix the notation:

\begin{equation}
{\cal L}_{eff}={\cal L}_{SM} + \sum_n \frac{\alpha_n}{\Lambda^2}{\cal O}^{(6)}_n + h.c.,
\end{equation}

\noindent where, ${\cal L}_{eff}$ is the effective Lagrangian which contains a series of higher-dimensional operators
built with the SM fields, ${\cal L}_{SM}$ is the renormalizable SM Lagrangian, $\Lambda$ is the mass scale at which new
physics expected to be observed, $\alpha_n$ are dimensionless coefficients and ${\cal O}^{(6)}_n$ represents the dimension-six
gauge-invariant operator.

\subsection{$\tau^+\tau^-\gamma$ vertex form factors}

The most general structure consistent with Lorentz and electromagnetic gauge invariant for the $\tau^+\tau^-\gamma$ vertex
describing the interaction of an on-shell photon $\gamma$ with two on-shell fermions $\tau^+\tau^-$ can be written in terms of
four form factors \cite{Passera1,Grifols,Escribano,Giunti,Giunti1}:

\begin{equation}
\Gamma^{\alpha}_\tau=eF_{1}(q^{2})\gamma^{\alpha}+\frac{ie}{2m_\tau}F_{2}(q^{2})\sigma^{\alpha\mu}q_{\mu}
+ \frac{e}{2m_\tau}F_3(q^2)\sigma^{\alpha\mu}q_\mu\gamma_5 +eF_4(q^2)\gamma_5(\gamma^\alpha - \frac{2q^\alpha m_\tau}{q^2}).
\end{equation}

\noindent In this expression, $q$ is the four-momentum of the photon, $e$ and $m_\tau$ are the charge of the electron and
the mass of the $\tau$-lepton. Since the two leptons are on-shell the form factors $F_{1,2,3,4}(q^2)$ appearing in Eq. (3) are
functions of $q^2$ and $m^2_\tau$ only, and have the following interpretations for $q^2=0$.

$i)$ $F_1(0)$ parameterize the vector part of the electromagnetic current and it is identified with the electric charge:

\begin{equation}
Q_\tau = F_1(0).
\end{equation}

$ii)$ $F_2(0)$ defines the anomalous MDM:

\begin{equation}
a_\tau = F_2(0).
\end{equation}

$iii)$ $F_3(0)$ describes the EDM:

\begin{equation}
d_\tau =\frac{e}{2m_\tau} F_3(0).
\end{equation}

$iv)$ $F_4(0)$ is the Anapole form factor:

\begin{equation}
F_A =-\frac{F_4}{q^2}.
\end{equation}

It is worth mentioning that in the SM at tree level, $F_1 = 1$ and $F_2 = F_3 = F_4=0$. In addition, should be noted that
the $F_2$ term behaves under C and P like the SM one, while the $F_3$ term violates CP.

\subsection{Gauge-invariant operators of dimension six for $\tau$-lepton dipole moments}

Theoretically, experimentally and phenomenologically most of the $\tau$-lepton anomalous electromagnetic vertices involve off-shell $\tau$-leptons.
In our study, one of the $\tau$-leptons is off-shell and measured quantity is not directly $a_\tau$ and $d_\tau$. For this reason deviations
of the $\tau$-lepton dipole moments from the SM values are examined in a model-independent way using the effective Lagrangian formalism.
This formalism is defined by high-dimensional operators which lead to anomalous $\tau^+ \tau^- \gamma$ coupling. For our study, we
apply the dimension-six effective operators that contribute to the MDM and EDM \cite{Buchmuller,1,eff1,eff3} of the $\tau$-lepton:

\begin{eqnarray}
{\cal L}_{eff}=\frac{1}{\Lambda^{2}} \Bigl[C_{LW}^{33} Q_{LW}^{33}+C_{LB}^{33} Q_{LB}^{33} + \mbox{h.c.}\Bigr],
\end{eqnarray}

\noindent where

\begin{eqnarray}
Q_{LW}^{33}&=&\bigl(\bar{\ell_{\tau}}\sigma^{\mu\nu}\tau_{R}\bigr)\sigma^{I}\varphi W_{\mu\nu}^{I},\\
Q_{LB}^{33}&=&\bigl(\bar{\ell_{\tau}}\sigma^{\mu\nu}\tau_{R}\bigr)\varphi B_{\mu\nu}.
\end{eqnarray}

\noindent Here $\ell_{\tau}$ is the tau leptonic doublet and $\varphi$ is the Higgs doublet, while $B_{\mu\nu}$ and $W_{\mu\nu}^{I}$
are the $U(1)_Y$ and $SU(2)_L$ gauge field strength tensors.

After electroweak symmetry breaking from the effective Lagrangian given by Eq. (8), the Higgs gets a vacuum expectation value
$\upsilon=246$ GeV and the corresponding CP even $\kappa$ and CP odd $\tilde{\kappa}$ observables are obtained:

\begin{eqnarray}
\kappa&=&\frac{2 m_{\tau}}{e} \frac{\sqrt{2}\upsilon}{\Lambda^{2}} Re\Bigl[\cos\theta _{W} C_{LB}^{33}- \sin\theta _{W} C_{LW}^{33}\Bigr],\\
\tilde{\kappa}&=&\frac{2 m_{\tau}}{e} \frac{\sqrt{2}\upsilon}{\Lambda^{2}} Im\Bigl[\cos\theta _{W} C_{LB}^{33}- \sin\theta _{W} C_{LW}^{33}\Bigr],
\end{eqnarray}

\noindent where, as usual $\sin\theta _{W} (\cos\theta _{W})$ is the sine (cosine) of the weak mixing angle.

The effective Lagrangian given by Eq. (8) gives additional contributions to the electromagnetic moments of the $\tau$-lepton,
which usually are expressed in terms of the parameters $\tilde a_\tau$ and $\tilde d_\tau$. They can be described in terms
of $\kappa$ and $\tilde{\kappa}$ as follows:

\begin{eqnarray}
\tilde{a}_{\tau}&=& \kappa,  \\
\tilde{d}_{\tau}&=&\frac{e}{2m_{\tau}}\tilde{\kappa}.
\end{eqnarray}

\section{The total cross-sections in $pp$, $e^-p$ and $e^+e^-$ colliders}

As we mentioned above, the $\tau$-lepton anomalous couplings offer an interesting window to physics BSM. Furthermore,
usually the current and future colliders probing the feasibility of measured the anomalous couplings that are enhanced for higher
values of the particle mass, making the $\tau$-lepton the ideal candidate among the leptons to observe these new couplings.

We point out that the total cross-section for the channels $pp \to p\tau\bar \tau \gamma p$ at the LHC, $e^-p \to e^- \tau\bar \tau \gamma p$
at the FCC-he and $e^+e^- \to e^+ \tau\bar \tau \gamma e^-$ at the CLIC are large enough to allow for a study of the anomalous
electromagnetic couplings of the $\tau$-lepton. The schematic diagram corresponding to these processes is given in Fig. 1, and
the subprocess $\gamma^*\gamma^* \to \tau\bar \tau \gamma$ can be produced via the set of Feynman diagrams depicted in Fig. 2.

It must be noticed that, unlike direct processes $e^+e^- \to \tau^+\tau^-$ \cite{ARGUS,Sher}, $e^+e^- \to \tau^+\tau^-\gamma$ \cite{L3},
$Z \to \tau^+\tau^-\gamma$ \cite{OPAL,Grifols} and $H \to \tau^+\tau^-\gamma$ \cite{Galon}, the two-photon processes
$\gamma^*\gamma^* \to \tau^+\tau^-\gamma$ offers several advantages to study the electromagnetic tau couplings at the LHC, FCC-he and CLIC.
The characteristics that distinguish them from the direct processes are mainly: 1) High sensitivity on $\tilde a_\tau$ and
$\tilde d_\tau$. 2) Increase of the cross-section for high energies and high luminosity. 3) They are extremely clean reactions because
there is no interference with weak interactions as they are purely quantum electrodynamics (QED) reactions. 4) The photon-photon fusion
processes are free from the uncertainties originated by possible anomalous $Z\gamma\gamma$ couplings. 5) Since the photons in the initial
state are almost real and the invariant mass of the tau-pairs is very small, we expect the effects of unknown form-factors to be negligible.
6) Furthermore, a very important feature is that the present and future colliders such as LHC, FCC-he and CLIC can produce very hard photons
at high luminosity in the Equivalent Photon Approximation (EPA) of high energy $pp$, $e^-p$ and $e^+e^-$ beams, with which the final state
photon identification has the advantage to determine the tau pair identification.

The main theoretical tool of our study for sensitivity estimates on the total cross-section of the processes
$pp \to p\tau\bar \tau \gamma p$, $e^-p \to e^- \tau\bar \tau \gamma p$ and $e^+e^- \to e^+ \tau\bar \tau \gamma e^-$
and on the anomalous $\tau$-lepton couplings, is the EPA. In the literature this approach is
commonly referred to as the Weizsacker-Williams Approximation (WWA) \cite{Weizsacker,Williams}. In general, EPA is a standard
semi-classical alternative to the Feynman rules for calculation of the electromagnetic interaction cross sections.
This approximation has many advantages. It helps to obtain crude numerical estimates through simple formulas. Furthermore,
this approach may principally ease the experimental analysis because it gives an opportunity one to directly achieve a rough
cross-section for $\gamma^*\gamma^* \to X$ subprocess through the research of the reaction $pp\hspace{0.8mm}(e^-p, e^+e^-)
\to pp\hspace{0.8mm} (e^-p, e^+e^-)X$, where $X$ symbolizes objects generated in the final state. The essence of the EPA is
as follows, photons emitted from incoming charged particles which have very low virtuality are scattered at very small angles
from the beam pipe and because the emitted quasi-real photons have a low $Q^2$ virtuality, these are almost real.

It is worth mentioning that the exclusive two-photon processes can be distinguished from fully inelastic processes by the following
experimental signatures: after of the elastic emission of a photon, incoming charged particles (electron or proton) are scattered with
a small angle and escapes detection from the central detectors. This generate a missing energy signature called forward large-rapidity
gap, in the corresponding forward region of the central detector \cite{Albrow}. This method have been observed experimentally at the LEP,
Tevatron and LHC \cite{Abulencia,Aaltonen1,Aaltonen2,Chatrchyan1,Chatrchyan2,Abazov,Chatrchyan3}.

Also, another experimental signature can be implemented by forward particle tagging. These detectors are to tag the electrons and protons
with some energy fraction loss. One of the well known applications of the forward detectors is the high energy photon induced interaction
with exclusive two lepton final states. Two almost real photons emitted by charged particles beams interact each other to produce two leptons
$\gamma^{*} \gamma^{*}\rightarrow \ell^{-} \ell^{+}$.  Deflected particles and their energy loss will be detected by the forward detectors
mentioned above but leptons will go to central detector. Produced lepton pairs have very small backgrounds \cite{Albrow2}. Use of very forward
detectors in conjunction with central detectors with a precise synchronization, can efficiently reduce backgrounds from pile-up events
\cite{Albrow1,Albrow2,Tasevsky,Tasevsky1}.

CMS and TOTEM Collaborations at the LHC began these measurements using forward detectors between the CMS interaction point and detectors in
the TOTEM area about $210\hspace{0.8mm}m$ away on both sides of interaction point \cite{Sirunyam}. However, LHeC and CLIC have a program of forward
physics with extra detectors located in a region between a few tens up to several hundreds of metres from the interaction point \cite{Fernandez,CLIC2018}.

\subsection{Benchmark parameters, selected cuts and $\chi^2$ fitting }

In this work, to evaluate the total cross-section $\sigma(pp \to p\tau\bar \tau \gamma p)$, $\sigma(e^-p \to e^- \tau\bar \tau \gamma p)$
and $\sigma(e^+e^- \to e^+ \tau\bar \tau \gamma e^-)$ and to probe the dipole moments $\tilde a_\tau$ and  $\tilde d_\tau$, we examine
the potential of LHC, FCC-he and CLIC based $\gamma^*\gamma^*$ colliders with the main parameters given in Table II. Furthermore, in order
to suppress the backgrounds and optimize the signal sensitivity, we impose for our study the following kinematic basic acceptance
cuts for $\tau^+ \tau^- \gamma$ events at the LHC, FCC-he and CLIC:

\begin{table}
\caption{Benchmark parameters of the LHC, FCC-he and CLIC based $\gamma^*\gamma^*$
colliders \cite{LHC0,LHC1,LHC2,LHC3,Fernandez,FCChe,Acar,Abramowicz}.}
\label{tab:1}
\begin{tabular}{|c|c|c|}
\hline
LHC            & $\sqrt{s}\hspace{0.8mm}(TeV)$      &  ${\cal L}(fb^{-1})$             \\
\hline\hline
Phase I     & 7, 8                                  & 10, 20, 30, 40, 50              \\
\hline
Phase II    & 13                                    & 10, 30, 50, 100, 200             \\
\hline
Phase III   & 14                                    & 10, 30, 50, 100, 200, 300, 3000     \\
\hline
\hline
FCC-he         &  $\sqrt{s}\hspace{0.8mm}(TeV)$    &  ${\cal L}(fb^{-1})$      \\
\hline\hline
Phase I       & 3.5                                & 20, 50, 100, 300, 500     \\
\hline
Phase II      & 7.07                               & 100, 300, 500, 700, 1000     \\
\hline
Phase III     & 10                                 & 100, 300, 500, 700, 1000     \\
\hline
\hline
CLIC           &  $\sqrt{s}\hspace{0.8mm}(TeV)$    &  ${\cal L}(fb^{-1})$      \\
\hline\hline
Phase I    & 0.350                             & 10, 50, 100, 200, 500     \\
\hline
Phase II   & 1.4                               & 10, 50, 100, 200, 500, 1000, 1500     \\
\hline
Phase III    & 3                                 & 10, 100, 500, 1000, 2000, 3000     \\
\hline
\end{tabular}
\end{table}

\begin{eqnarray}
\begin{array}{c}
p^\gamma_t > 20 \hspace{0.8mm}GeV, \hspace{5mm} |\eta^{\gamma}|< 2.5,\\
p^{\tau^+, \tau^-}_t > 20\hspace{0.8mm}GeV,    \hspace{5mm} |\eta^{\tau^+, \tau^-}|< 2.5,\\
\Delta R(\tau^-, \gamma) >0.4,\\
\Delta R(\tau^+, \tau^-) >0.4,\\
\Delta R(\tau^+, \gamma) >0.4.\\
\end{array}
\end{eqnarray}

\noindent Here the cuts given by Eq. (15) are applied to the photon transverse momentum $p^\gamma_t$, to the photon pseudorapidity
$\eta^\gamma$, which reduces the contamination from other particles misidentified as photons, to the tau transverse momentum
$p^{\tau^-, \tau^+}_t$ for the final state particles, to the tau pseudorapidity $\eta^\tau$ which reduces the contamination from
other particles misidentified as tau and to $\Delta R(\tau^-, \gamma)$, $\Delta R(\tau^-, \tau^+)$ and $\Delta R(\tau^+, \gamma)$
which give the separation of the final state particles. It is fundamental that we apply these cuts to reduce the background and to
optimize the signal sensitivity to the particles of the $\tau^+\tau^-\gamma$ final state.

Tau identification efficiency depends of a specific process, background processes, some kinematic parameters and luminosity. For the
processes examined, investigations of tau identification have not been examined yet for LHC, FCC-he and CLIC detectors. In this case,
identification efficiency can be detected as a function of transverse momentum and rapidity of the $\tau$-lepton. We have considered
the following cuts for the selection of the $\tau$-lepton as used in many studies \cite{Galon,Atag2}
$p_{t}^{\tau^{+},\tau^{-}} > 20\hspace{0.8mm}GeV$, $\vert \eta^{\tau^{+},\tau^{-}} \vert  < 2.5$.

The above cuts on the $\tau$-leptons ensure that their decay products are collimated which allows their momenta to be reconstructed reasonably
accurately, despite the unmeasured energy going into neutrinos \cite{Howard}.

Another important element in our study is the level or degree of sensitivity of our results. In this sense, to estimate the $95\%$
Confidence Level (C.L.) sensitivity on the parameters ${\tilde a}_\tau$ and ${\tilde d}_\tau$, a $\chi^2$ fitting is performed.
The $\chi^2$ distribution \cite{murat,Billur} is defined by

\begin{equation}
\chi^2(\tilde a_\tau, \tilde d_\tau)=\Biggl(\frac{\sigma_{SM}-\sigma_{BSM}(\sqrt{s}, \tilde a_\tau, \tilde d_\tau)}{\sigma_{SM}\sqrt{(\delta_{st})^2
+(\delta_{sys})^2}}\Biggr)^2,
\end{equation}

\noindent with $\sigma_{BSM}(\sqrt{s}, \tilde a_\tau, \tilde d_\tau)$ is the total cross-section incorporating contributions from the SM
and new physics, $\delta_{st}=\frac{1}{\sqrt{N_{SM}}}$ is the statistical error and $\delta_{sys}$ is the systematic error. The number
of events is given by $N_{SM}={\cal L}_{int}\times \sigma_{SM}\times BR$, where ${\cal L}_{int}$ is the integrated luminosity of the
$pp$, $e^-p$ and $e^+e^-$ colliders. The $\tau$-lepton decays almost $17.8\%$ of the time into an electron and into two neutrinos, $17.4\%$
of the time, it decays in a muon and in two neutrinos. While, in the remaining $64.8\%$ of the occasions, it decays in the form of hadrons
and a neutrino. Thus, we assume that the branching ratio of the $\tau$-lepton pair in the final state to be $BR(\mbox{Pure-leptonic}) = 0.123$
or $BR(\mbox{Semi-leptonic}) = 0.46$ \cite{Data2018}.

On the other hand, it should be noted that in all the processes considered in this article, the total cross-section of the
$pp \to p\tau\bar \tau \gamma p$, $e^-p \to e^- \tau\bar \tau \gamma p$ and $e^+e^- \to e^+ \tau\bar \tau \gamma e^-$ signals
are computed using the CalcHEP package \cite{Belyaev}, which can computate the Feynman diagrams, integrate over multiparticle
phase space, and simulate events.

\subsection{The total cross-section of the $pp \to p\gamma^*\gamma^*p \to p \tau^+\tau^-\gamma p$ signal at LHC}

In the EPA, the quasireal photons emitted from both proton beams collide with each other and produce the subprocess
$\gamma^{*} \gamma^{*} \rightarrow \tau^- \tau^+ \gamma$. The spectrum of photon emitted by proton can be written
as follows \cite{Belyaev,Budnev}:

\begin{eqnarray}
f_{\gamma^{*}_p}(x)=\frac{\alpha}{\pi E_{p}}\{[1-x][\varphi(\frac{Q_{max}^{2}}{Q_{0}^{2}})-\varphi(\frac{Q_{min}^{2}}{Q_{0}^{2}})],
\end{eqnarray}

\noindent where $x=E_{\gamma^*_p}/E_p$ and $Q^2_{max}$ is maximum virtuality of the photon. The minimum value of the $Q^2_{min}$ is given by

\begin{eqnarray}
Q^2_{min}=\frac{m_p^2x^2}{1-x}.
\end{eqnarray}

\noindent The function $\varphi$ is given by

\begin{eqnarray}
\varphi(\theta)=&&(1+ay)\left[-\textit{In}(1+\frac{1}{\theta})+\sum_{k=1}^{3}\frac{1}{k(1+\theta)^{k}}\right]
+\frac{y(1-b)}{4\theta(1+\theta)^{3}} \nonumber \\
&& +c(1+\frac{y}{4})\left[\textit{In}\left(\frac{1-b+\theta}{1+\theta}\right)+\sum_{k=1}^{3}\frac{b^{k}}{k(1+\theta)^{k}}\right].
\end{eqnarray}

\noindent with

\begin{eqnarray}
y=\frac{x^{2}}{(1-x)},
\end{eqnarray}

\begin{eqnarray}
a=\frac{1+\mu_{p}^{2}}{4}+\frac{4m_{p}^{2}}{Q_{0}^{2}}\approx 7.16,
\end{eqnarray}

\begin{eqnarray}
b=1-\frac{4m_{p}^{2}}{Q_{0}^{2}}\approx -3.96,
\end{eqnarray}

\begin{eqnarray}
c=\frac{\mu_{p}^{2}-1}{b^{4}}\approx 0.028.
\end{eqnarray}

Therefore, in the EPA the total cross-section of the $pp \to p\gamma^*\gamma^*p \to p \tau^+\tau^-\gamma p$ signal is given by

\begin{eqnarray}
\sigma_{pp \to p\gamma^*\gamma^*p \to p\tau^+\tau^-\gamma p}=\int f_{\gamma^*_p}(x)f_{\gamma^*_p}(x)d{\hat\sigma}_{\gamma^*\gamma^* \to \tau^+\tau^-\gamma}dE_{1}dE_{2}.
\end{eqnarray}

With all the elements considered in subsection A, that is to say the CalcHEP package, selected cuts, $\chi^2$ fitting and with
13 and 14 TeV at the LHC, the determination of the total cross-section in terms of the anomalous parameters $\kappa$ and $\tilde\kappa$,
translate in the following results: \\

$i)$ For $\sqrt{s}=13\hspace{0.8mm} TeV$:

\begin{eqnarray}
\sigma(\kappa)&=&\Bigl[1.28\times 10^7 \kappa^6 + 2.61\times10^3 \kappa^5 + 4.71\times10^3 \kappa^4 + 1.99 \kappa^3  \nonumber\\
&+& 1.39 \kappa^2 + 2.34\times10^{-4}\kappa + 1.03\times10^{-4} \Bigr] (pb),   \\
\sigma(\tilde{\kappa})&=&\Bigl[1.28\times10^7\tilde{\kappa}^6 + 4.71\times10^3\tilde{\kappa}^4 + 1.39\tilde\kappa^2 + 1.03\times10^{-4} \Bigr] (pb).
\end{eqnarray}

$ii)$ For $\sqrt{s}=14\hspace{0.8mm} TeV$:

\begin{eqnarray}
\sigma(\kappa)&=&\Bigl[1.78\times 10^7 \kappa^6 + 3.52\times10^3 \kappa^5 + 5.18\times10^3 \kappa^4 + 2.21 \kappa^3  \nonumber\\
&+& 1.50 \kappa^2 + 2.40\times10^{-4}\kappa + 1.05\times10^{-4} \Bigr] (pb),   \\
\sigma(\tilde{\kappa})&=&\Bigl[1.78\times10^7\tilde{\kappa}^6 + 5.18\times10^3\tilde{\kappa}^4 + 1.50\tilde\kappa^2 + 1.05\times10^{-4} \Bigr] (pb).
\end{eqnarray}

In these expressions the independent terms of $\kappa$ and $\tilde\kappa$ correspond to the cross-section of the SM, that is
$\kappa=\tilde\kappa=0$. In the next section, the calculated cross-sections in Eqs. (25)-(28) are used to sensitivity estimates
on the anomalous MDM and EDM of the $\tau$-lepton.

\subsection{The total cross-section of the $e^-p \to e^-\gamma^*\gamma^*p \to e^- \tau^+\tau^-\gamma p$ signal at FCC-he}

To determine the total cross-section of the $e^-p \to e^-\gamma^*\gamma^*p \to e^- \tau^+\tau^-\gamma p$ signal at FCC-he,
we must take into account that in the EPA approach, the spectrum of first photon emitted by electron is given by \cite{Belyaev,Budnev}:

\begin{eqnarray}
f_{\gamma^{*}_e}(x_{1})=\frac{\alpha}{\pi E_{e}}\{[\frac{1-x_{1}+x_{1}^{2}/2}{x_{1}}]log(\frac{Q_{max}^{2}}{Q_{min}^{2}})-\frac{m_{e}^{2}x_{1}}{Q_{min}^{2}}
&&(1-\frac{Q_{min}^{2}}{Q_{max}^{2}})-\frac{1}{x_{1}}[1-\frac{x_{1}}{2}]^{2}log(\frac{x_{1}^{2}E_{e}^{2}+Q_{max}^{2}}{x_{1}^{2}E_{e}^{2}+Q_{min}^{2}})\}, \nonumber \\
\end{eqnarray}

\noindent where $x_1=E_{\gamma^*_e}/E_{e}$ and $Q^2_{max}$ is maximum virtuality of the photon. The minimum value of the $Q^2_{min}$
is given by

\begin{eqnarray}
Q_{min}^{2}=\frac{m_{e}^{2}x_{1}^{2}}{1-x_{1}}.
\end{eqnarray}

For the spectrum $f_{\gamma^{*}_p}(x_2)$ of the second photon emitted by proton we consider the expression given by Eq. (17).
Therefore, the total cross-section of the reaction $e^-p \to e^-\gamma^*\gamma^*p \to e^- \tau^+\tau^-\gamma p$ is obtained from

\begin{eqnarray}
\sigma_{e^-p \rightarrow e^-\gamma^{*} \gamma^{*} p\rightarrow e^- \tau^+ \tau^- \gamma p}=\int f_{\gamma^{*}_e}(x_{1})f_{\gamma^{*}_p}(x_{2}) d\hat{\sigma}_{\gamma^{*} \gamma^{*} \rightarrow \tau^+ \tau^-\gamma} dx_{1} dx_{2}.
\end{eqnarray}

We have performed a global fit (and apply the cuts given in Eq. (15)), as a function of the two independent anomalous couplings
$\kappa$ and $\tilde{\kappa}$, with 7.07 and  10 TeV at FCC-he to the following studied observables:\\

$i)$ For $\sqrt{s}=7.07\hspace{0.8mm} TeV$:

\begin{eqnarray}
\sigma(\kappa)&=&\Bigl[2.85\times 10^7 \kappa^6 + 2.33\times10^3 \kappa^5 + 1.87\times10^4 \kappa^4 + 11.14 \kappa^3  \nonumber\\
&+& 7.30 \kappa^2 + 1.65\times10^{-3}\kappa + 6.09\times10^{-4} \Bigr] (pb),   \\
\sigma(\tilde{\kappa})&=&\Bigl[2.85\times10^7\tilde{\kappa}^6 + 1.87\times10^4\tilde{\kappa}^4 + 7.30\tilde\kappa^2 + 6.09\times10^{-4} \Bigr] (pb).
\end{eqnarray}

\newpage

$ii)$ For $\sqrt{s}=10\hspace{0.8mm} TeV$:

\begin{eqnarray}
\sigma(\kappa)&=&\Bigl[2.10\times 10^8 \kappa^6 + 3.26\times10^4 \kappa^5 + 6.40\times10^4 \kappa^4 + 13.85 \kappa^3  \nonumber\\
&+& 12.11 \kappa^2 + 2.46\times10^{-3}\kappa + 8.50\times10^{-4} \Bigr] (pb),   \\
\sigma(\tilde{\kappa})&=&\Bigl[2.10\times10^8\tilde{\kappa}^6 + 6.40\times10^4\tilde{\kappa}^4 + 12.11\tilde\kappa^2 + 8.50\times10^{-4} \Bigr] (pb).
\end{eqnarray}

\subsection{The total cross-section of the $e^+e^- \to e^+ \gamma^*\gamma^* e^- \to e^+ \tau^+\tau^-\gamma e^-$ signal at CLIC}

The total cross-section for the elementary $e^+e^- \to e^+ \gamma^*\gamma^* e^- \to e^+ \tau^+\tau^-\gamma e^-$ processes at CLIC
is determined in the context of EPA, where the quasi-real photons emitted from both lepton beams collide with each other
and produce the subprocess $\gamma^{*} \gamma^{*} \rightarrow \tau^+ \tau^-\gamma$.

The form of the spectrum in two-photon collision energy $f_{\gamma^{*}}(x)$ is a very important ingredient in the EPA.
In this approach, the photon energy spectrum is given by Eqs. (29) and (30).

The elementary $\gamma^{*} \gamma^{*} \rightarrow \tau^+ \tau^- \gamma$ process participates as a subprocess in the main process
$e^+e^- \to e^+ \gamma^*\gamma^* e^- \to e^+ \tau^+\tau^-\gamma e^-$, and the total cross-section is given by

\begin{eqnarray}
\sigma_{e^+e^- \to e^+\gamma^*\gamma^* \to e^+\tau^+\tau^-\gamma e^-}=\int f_{\gamma^*_{e^-}}(x)f_{\gamma^*_{e^+}}(x)d\hat{\sigma}_{\gamma^*\gamma^* \to \tau^+\tau^-\gamma}dE_{1}dE_{2}.
\end{eqnarray}

We presented results for the dependence of the total cross-section of the process $\gamma^{*} \gamma^{*} \rightarrow \tau^+ \tau^- \gamma$
on $\kappa$ and $\tilde\kappa$. We consider the following cases at CLIC:\\

$i)$ For $\sqrt{s}=1.5\hspace{0.8mm} TeV$:

\begin{eqnarray}
\sigma(\kappa)&=&\Bigl[2.09\times 10^7 \kappa^6 + 5.09\times10^4 \kappa^5 + 5.86\times10^4 \kappa^4 + 63.89 \kappa^3 + 60 \kappa^2  \nonumber\\
&+& 2.10\times10^{-2}\kappa + 6.9\times10^{-3} \Bigr] (pb)   \\
\sigma(\tilde{\kappa})&=&\Bigl[2.09\times10^7\tilde{\kappa}^6 + 5.86\times10^4\tilde{\kappa}^4 + 60\tilde{\kappa}^2
+ 6.9\times10^{-3} \Bigr] (pb).
\end{eqnarray}

\newpage

$ii)$ For $\sqrt{s}=3\hspace{0.8mm} TeV$:

\begin{eqnarray}
\sigma(\kappa)&=&\Bigl[4.22\times 10^8 \kappa^6 + 2.49\times10^5 \kappa^5 + 2.89\times10^5 \kappa^4 + 1.25\times10^2 \kappa^3  \nonumber\\
&+& 1.22\times10^2 \kappa^2 + 2.79\times10^{-2}\kappa + 9.74\times10^{-3} \Bigr] (pb)   \\
\sigma(\tilde{\kappa})&=&\Bigl[4.22\times10^8\tilde{\kappa}^6 + 2.89\times10^5\tilde{\kappa}^4 + 1.22\times10^2 \tilde\kappa^2 + 9.74\times10^{-3} \Bigr] (pb).
\end{eqnarray}

In the next section, the calculated cross-sections in Eqs. (25)-(28), (32)-(35) and (37)-(40) are used to sensibility estimates
on the anomalous MDM and EDM of the $\tau$-lepton.

\section{Sensitivity estimates on the dipole moments of the $\tau$-lepton at the LHC, FCC-he and CLIC}

\subsection{Sensibility on the dipole moments of the $\tau$-lepton from $pp \to p\tau^+ \tau^- \gamma p$ at LHC}

In this subsection phenomenological projections on the total cross-section and on the dipolar moments $\kappa$ and $\tilde\kappa$ of
the $\tau$-lepton though the $pp \to p\tau^+ \tau^- \gamma p$ signal at LHC are presented.

For our numerical analysis we starting from the expressions given by Eqs. (25)-(28) and we obtained the total cross-sections plots
of Figs. 3-6. These four figures represent the same observable, but just expressed in terms of different anomalous parameters,
that is $\kappa$, $\tilde\kappa$ and $(\kappa, \tilde\kappa)$, respectively. From these figures, a strong dependence of the total
cross-section with respect to the anomalous parameters $\kappa$, $\tilde\kappa$, as well as with the center-of-mass energies of
the LHC is clearly observed. Furthermore, a direct comparison between the results for the SM, that is to say with $\kappa=\tilde\kappa=0$
(see Eqs. (25)-(28)) and the corresponding ones obtained in Figs. 3-6, show a great difference of the order of ${\cal O}(10^3-10^4)$
on the total cross-section.

To estimate the sensitivity of the LHC to the anomalous couplings $\kappa$ and $\tilde\kappa$ we consider $\sqrt{s}=13$ and 14 TeV
and integrated luminosities ${\cal L}= 10, 50, 200\hspace{0.8mm}fb^{-1}$. To this effect, in Figs. 7 and 8, we use Eq. (25)-(28) to
illustrate the region of parameter space allowed at $95\%$ C.L.. The best sensitivity estimated from Figs. 7 and 8, taken one coupling
at a time are given by:

\begin{equation}
\begin{array}{ll}
\kappa=(-0.007, 0.007), \hspace{3mm}  & \mbox{$95\%$ C.L.}, \\
\tilde\kappa=(-0.008, 0.008), \hspace{3mm}  & \mbox{$95\%$ C.L.},
\end{array}
\end{equation}

\noindent at $\sqrt{s}=14\hspace{0.8mm}TeV$ and ${\cal L}= 200\hspace{0.8mm}fb^{-1}$. These results are consistent with
those reported in Table III for $\sqrt{s}$ and ${\cal L}$ as in Eq. (42):

\begin{equation}
\begin{array}{ll}
\tilde a_\tau=(-0.0067, 0.0065), \hspace{3mm}  & \mbox{$95\%$ C.L.}, \\
|\tilde d_\tau|=3.692\times 10^{-17}\hspace{0.8mm}{\rm ecm}, \hspace{3mm}  & \mbox{$95\%$ C.L.}.
\end{array}
\end{equation}

Our results are an order of magnitude better than the best existing limit for the $\tau$-lepton anomalous MDM and EDM comes from the
process $e^+e^- \to e^+e^-\tau^+\tau^-$ as measured by DELPHI Collaboration \cite{DELPHI} at LEP2 (see Table I), as well as of
the study of $e^+e^- \to \tau^+\tau^-$  by BELLE Collaboration \cite{BELLE} (see Table I).

We next consider the sensibility estimated for the anomalous observables $\tilde a_\tau$ and $\tilde d_\tau$, considering different
values of $\sqrt{s}$ and ${\cal L}$ at $95\%$ C.L.. We consider both cases: pure-leptonic and semi-leptonic. Our results for these
cases are shown in Table III, where the semi-leptonic case provides more sensitive results on $\tilde a_\tau$ and $\tilde d_\tau$.

\subsection{Sensitivity on the dipole moments of the $\tau$-lepton from $e^-p \to e^- \tau\bar \tau \gamma p$ at FCC-he}

We now turn our attention to the associated production of a photon with a $\tau$-lepton pair, via the
$e^-p \to e^-\tau^+ \tau^- \gamma p$ signal, as is show in Figs. 9-12. The motivation to study this process is simple
and already mentioned above, the gauge invariance of the effective Lagrangian relates the dipole couplings of the $\tau$-lepton
to couplings involving the photon. At the same time a similar study of the total cross-section as a function of $\tau$-lepton
dipole couplings $\kappa$ and $\tilde\kappa$ are realized. Our results show that the total cross-section depends significantly
on $\kappa$ and $\tilde\kappa$, in addition to $\sqrt{s}$. We find that the difference with respect to the SM is of the order of
${\cal O}(10^3-10^5)$, which is several orders of magnitude best than the result of the SM.

Figs. 13 and 14 show the sensitivity contour bands in the plane of $\tilde\kappa$ vs $\kappa$ for the FCC-he with
center-of-mass energies $\sqrt{s}=7.07, 10 \hspace{0.8mm}TeV$ and luminosities ${\cal L}=100, 500, 1000\hspace{0.8mm}fb^{-1}$.
The sensitivity estimates at $95\%$ C.L. on the anomalous parameters are found to be:

\begin{equation}
\begin{array}{ll}
\kappa=(-0.0035, 0.0025), \hspace{3mm}  & \mbox{$95\%$ C.L.}, \\
\tilde\kappa=(-0.0025, 0.0030), \hspace{3mm}  & \mbox{$95\%$ C.L.}.
\end{array}
\end{equation}

\noindent Here, it was studied using data collected by the DELPHI experiment at LEP2 during the years 1997-2000. The corresponding
integrated luminosity is $650\hspace{0.8mm}pb^{-1}$. However, the corresponding integrated luminosity related to BELLE is
$29.5\hspace{0.8mm}fb^{-1}$.

The comparison with the limits of the present DELPHI and BELLE Collaborations with the corresponding ones obtained by the FCC-he on
the anomalous couplings searches, indicates that the sensitivity estimates of the FCC-he at $95\%$ C.L are still stronger that for
both experiments.

In Table IV, we list the $95\%$ C.L. sensitivity estimates on the observables $\tilde a_\tau$ and $\tilde d_\tau$, based on
di-tau production cross-section via the process $e^-p \to e^- \tau\bar \tau \gamma p$ at FCC-he. At present, DELPHI and BELLE
experimental measurements on tau pair production $e^+e^-\tau^+\tau^-$ and $\tau^+\tau^-$ give the most stringent bounds on
$\tilde a_\tau$ and $\tilde d_\tau$ \cite{DELPHI,BELLE}. However, note that our sensitivity estimates on $\tilde a_\tau$ and
$\tilde d_\tau$ are about ten times better than those for DELPHI and BELLE Collaborations, corroborating the impact of the
$e^-p \to e^- \tau\bar \tau \gamma p$ signal, in addition of the parameters of the FCC-he:

\begin{equation}
\begin{array}{ll}
\tilde a_\tau=(-0.00265, 0.00246), \hspace{3mm}  & \mbox{$95\%$ C.L.}, \\
|\tilde d_\tau|=1.437\times 10^{-17}\hspace{0.8mm}{\rm ecm}, \hspace{3mm}  & \mbox{$95\%$ C.L.}.
\end{array}
\end{equation}

\noindent with $\sqrt{s}=10\hspace{0.8mm}TeV$ and ${\cal L}= 1000\hspace{0.8mm}fb^{-1}$.

\subsection{Sensitivity on the dipole moments of the $\tau$-lepton from $e^+e^- \to e^+ \tau\bar \tau \gamma e^-$ at CLIC}

Before beginning with the study of the sensitivity on the dipole moments of the $\tau$-lepton through the process
$e^+e^- \to e^+ \tau\bar \tau \gamma e^-$ at CLIC, it should be noted that experimentally, the processes that involving single-photon
in the final state $\tau^{+}\tau^{-}\gamma$ can potentially distinguish from background associated with the process under consideration.
Besides, the anomalous $\tau^{+}\tau^{-}\gamma$ coupling can be analyzed through the process $e^{+}e^{-} \rightarrow \tau^{+}\tau^{-}$
at the linear colliders. This process receives contributions from both anomalous $\tau^{+}\tau^{-}\gamma$ and $\tau^{+}\tau^{-} Z$ couplings.
But, the subprocess $\gamma^* \gamma^* \rightarrow \tau^{+} \tau^{-} \gamma$ isolate $\tau^{+}\tau^{-} \gamma$ coupling which provides the
possibility to analyze the $\tau^{+}\tau^{-} \gamma$ coupling separately from the $\tau^{+}\tau^{-} Z$ coupling. Generally, anomalous parameters
$\tilde a_\tau$ and $\tilde d_\tau$ tend to increase the cross-section for the subprocess $\gamma^*\gamma^* \rightarrow \tau^{+}\tau^{-} \gamma $,
especially for photons with high energy which are well isolated from the decay products of the taus \cite{L3}. Furthermore, the single-photon
in the final state has the advantage of being identified with high efficiency and purity.

To assess future CLIC sensitivity to the dipole moments, as well as for the total cross-section from searches for the
$e^+e^- \to e^+ \tau\bar \tau \gamma e^-$ signal, we perform several figures, as well as a table that illustrates the
sensitivity on the dipole moments.

In Figs. 15-18 we show the expected $\sigma$ vs $\kappa$, $\sigma$ vs $\tilde\kappa$ and $\sigma$ vs $(\kappa, \tilde\kappa)$
cross-sections for the signal with $\sqrt{s}=1.5, 3 \hspace{0.8mm}TeV$. All analysis cuts given in Eq. (15) are applied. Obviously
of the plots we observed that the cross-section depends strongly on $\kappa$, $\tilde\kappa$ and $(\tilde\kappa, \kappa)$,
throughout the range defined for these observables, as well as of $\sqrt{s}$. An improvement of the order of ${\cal O}(10^3-10^2)$
with respect to the SM is obtained.

In Figs. 19 and 20, we show the exclusion contours on the two-parameter $\kappa$ and $\tilde\kappa$. For comparison, we also include
several energies and luminosities. The results of the CLIC improve the sensitivity of the existing limits for the MDM and the EDM of
the $\tau$-lepton given in Table I. What is more, there is also a significant improvement in the cross-section constraining $\kappa$
and $\tilde\kappa$ parameters because our observables are sensitive to the parameters of the collider. Thus, from Fig. 20, it is
straightforward to obtain that the sensitivity estimates on the anomalous dipole moment are:

\begin{equation}
\begin{array}{ll}
\kappa=(-0.00125, 0.00115), \hspace{3mm}  & \mbox{$95\%$ C.L.}, \\
\tilde\kappa=(-0.00125, 0.00125), \hspace{3mm}  & \mbox{$95\%$ C.L.},
\end{array}
\end{equation}

\noindent where the results obtained in Eq. (45) are for $\sqrt{s}=3\hspace{0.8mm}TeV$ and ${\cal L}=3000\hspace{0.8mm}fb^{-1}$.

Our final results are summarised in Table V below and agree with the experimental determinations of the $\tau$-lepton dipole
moments which are given in Table I for the DELPHI, L3, OPAL, BELLE and ARGUS Collaborations. From Table V, our best sensitivity
projected correspond to:

\begin{equation}
\begin{array}{ll}
\tilde a_\tau=(-0.00128, 0.00105), \hspace{3mm}  & \mbox{$95\%$ C.L.}, \\
|\tilde d_\tau|=6.439\times 10^{-18}\hspace{0.8mm}{\rm ecm}, \hspace{3mm}  & \mbox{$95\%$ C.L.}.,
\end{array}
\end{equation}

\noindent and the results obtained in Eq. (46) are for $\sqrt{s}=3\hspace{0.8mm}TeV$ and ${\cal L}=3000\hspace{0.8mm}fb^{-1}$ at CLIC.

It is worth mentioning that, the above sensitivity estimates are completely model-independent and no assumption has been made on
the anomalous couplings in the effective Lagrangian given by (8). For the sake of comparison with published data for the DELPHI,
L3, OPAL, BELLE and ARGUS Collaborations \cite{DELPHI,L3,OPAL,BELLE,ARGUS}, we have presented the limits that can be found by
considering separately only operator $\tilde a_\tau$ or only operator $\tilde d_\tau$ in Eqs. (8).

\section{Conclusions}

The sensitivity estimates of the $\tau^+\tau^-\gamma$ vertex at the LHC, FCC-he and CLIC at CERN are discussed in this paper.
We propose to measure this vertex in the $pp \to p\tau\bar \tau \gamma p$, $e^-p \to e^- \tau\bar \tau \gamma p$ and
$e^+e^- \to e^+ \tau\bar \tau \gamma e^-$ channels at the $\gamma^*\gamma^* \to \tau^+\tau^-\gamma$ mode. Furthermore, to the
total cross-section measurement with the EPA, $\chi^2$  method provides powerful tools to probe the anomalous structure of the
$\tau^+\tau^-\gamma$ coupling. Additionally, in order to select the events we implementing the standard isolation cuts,
compatibly with the detector resolution expected at LHC, FCC-he and CLIC to reduce the background and to optimize the
signal sensitivity.

A very important aspect in our study and worth mentioning is the following, in most of the above mentioned experiments some of the
particles in the anomalous $\tau^+\tau^-\gamma$ coupling are off-shell. The off-shell form factors are problematic
since they can hardly be isolated from other contributions and gauge invariance can be a difficulty. However, in the effective Lagrangian
approach which we use in this paper, all those difficulties are solved because form factors are directly related to couplings, which are
gauge invariant. Therefore, stringent and clean sensitivity estimates on the anomalous MDM and EDM of the $\tau$-lepton are obtained.

In conclusion, new mechanism are proposed in this paper obtain the anomalous MDM $(\tilde a_\tau)$ and EDM $(\tilde d_\tau)$ of the
$\tau$-lepton produced in the high energy $pp$, $e^-p$ and $e^+e^-$ colliders. With the information from the effective Lagrangian formalism,
a significant improvement can be achieved as shown in Figs. 3-20 and Tables III-V. Under this framework, it is predicted that with
$200\hspace{0.8mm}fb^{-1}$ of data that will be collected by LHC, a sensitivity of $\tilde a_\tau=(-0.0067, 0.0065)$, $95\%$ C.L., can be
achieved for the $\tau$-lepton, and $|\tilde d_\tau|=3.692\times 10^{-17}\hspace{0.8mm}{\rm ecm}$, $95\%$ C.L. can be achieved for the EDM.
In the case of the FCC-he, it is feasibility that with $1000\hspace{0.8mm}fb^{-1}$ it is possible to obtain a sensitivity of
$\tilde a_\tau=(0.00265, 0.00246)$ and $|\tilde d_\tau|=1.437\times 10^{-17}\hspace{0.8mm}{\rm ecm}$, at $95\%$ C.L.. While for the CLIC,
the projections with $3000\hspace{0.8mm}fb^{-1}$ of data that will be collected by CLIC are $\tilde a_\tau=(0.00128, 0.00105)$ and
$|\tilde d_\tau|=6.439\times 10^{-18}\hspace{0.8mm}{\rm ecm}$, at $95\%$ C.L.. The precision of the $\tau$-lepton is about $39\%$ of the SM
prediction, therefore in this framework and with the large amount of data collected at current and future colliders can constrain
BSM much better than before. In summary, the future CLIC at high energy and high luminosity should provide the best sensitivity
on the MDM and EDM of the $\tau$-lepton, and shows a potential advantage compared to those from LHC and FCC-he.

\begin{table}[!ht]
\caption{Model-independent sensitivity estimate for the $\tilde{a}_\tau$ magnetic moment and the $\tilde{d}_\tau$ electric dipole
moment through the process $p\,p \to p\,\tau^+ \tau^-\, \gamma \, p$ at LHC.}
\begin{center}
\begin{tabular}{|c|cc|cc|}
\hline\hline
\multicolumn{5}{|c|}{ $\sqrt{s}= 13$ TeV,  \hspace{1cm}  $95\%$ C.L.}\\
 \hline
 \cline{1-5}
 & \multicolumn{2}{|c|}{Pure-leptonic} & \multicolumn{2}{|c|}{Semi-leptonic}\\
 \hline
 ${\cal L}\,(fb^{-1})$  & $\tilde a_\tau$ & \hspace{1.8cm}$ |\tilde{d}_\tau({\rm ecm})|$\hspace{1cm} & $\tilde a_\tau$  & \hspace{1.8cm}$
 |\tilde{d}_\tau({\rm ecm})|$ \hspace{1cm} \\
\hline
10   & [-0.02051, 0.02038]    & 1.1361$\times 10^{-16}$  & [-0.01406, 0.01391]      & 7.7757$\times 10^{-17}$  \\
30   & [-0.01514, 0.01499]    & 8.3785$\times 10^{-17}$  & [-0.01077, 0.01061]      & 5.9451$\times 10^{-17}$ \\
50   & [-0.01488, 0.01473]    & 8.2331$\times 10^{-17}$  & [-0.00952, 0.00935]      & 5.2471$\times 10^{-17}$  \\
100  & [-0.01139, 0.01122]    & 6.2865$\times 10^{-17}$  & [-0.00845, 0.00828]      & 4.6533$\times 10^{-17}$  \\
200  & [-0.01030, 0.01014]    & 5.6832$\times 10^{-17}$  & [-0.00708, 0.00691]      & 3.8904$\times 10^{-17}$ \\
\hline
 \cline{1-5}
\multicolumn{5}{|c|}{ $\sqrt{s}= 14$ TeV,  \hspace{1cm}  $95\%$ C.L.}\\
\hline
10   & [-0.01959, 0.01948]   & 1.0860$\times 10^{-16}$  & [-0.01346, 0.01333]      & 7.4504$\times 10^{-17}$  \\
30   & [-0.01449, 0.01436]   & 8.0268$\times 10^{-17}$  & [-0.01031, 0.01016]      & 5.6889$\times 10^{-17}$ \\
50   & [-0.01424, 0.01411]   & 7.8880$\times 10^{-17}$  & [-0.00911, 0.00895]      & 5.0127$\times 10^{-17}$  \\
100  & [-0.01090, 0.01076]   & 6.0188$\times 10^{-17}$  & [-0.00807, 0.00792]      & 4.4358$\times 10^{-17}$  \\
200  & [-0.00986, 0.00971]   & 5.4354$\times 10^{-17}$  & [-0.00674, 0.00658]      & 3.6925$\times 10^{-17}$ \\
\hline\hline
\end{tabular}
\end{center}
\end{table}

\begin{table}[!ht]
\caption{Model-independent sensitivity estimate for the $\tilde{a}_\tau$ magnetic moment and the $\tilde{d}_\tau$ electric dipole
moment through the process $e^- p \to e^-\,\tau^+ \tau^-\, \gamma \, p$ at FCC-he.}
\begin{center}
\begin{tabular}{|c|cc|cc|}
\hline\hline
\multicolumn{5}{|c|}{ $\sqrt{s}= 7.07$ TeV,  \hspace{1cm}  $95\%$ C.L.}\\
 \hline
 \cline{1-5}
 & \multicolumn{2}{|c|}{Pure-leptonic} & \multicolumn{2}{|c|}{Semi-leptonic}\\
 \hline
 ${\cal L}\,(fb^{-1})$  & $\tilde a_\tau$ & \hspace{1.8cm}$ |\tilde{d}_\tau({\rm ecm})|$\hspace{1cm} & $\tilde a_\tau$  & \hspace{1.8cm}$
 |\tilde{d}_\tau({\rm ecm})|$ \hspace{1cm} \\
\hline
100   & [-0.00752, 0.00728]   & 4.1119$\times 10^{-17}$  & [-0.00548, 0.00525]    & 2.9760$\times 10^{-17}$  \\
300   & [-0.00576, 0.00553]   & 3.1321$\times 10^{-17}$  & [-0.00425, 0.00402]    & 2.2956$\times 10^{-17}$ \\
500   & [-0.00513, 0.00490]   & 2.7808$\times 10^{-17}$  & [-0.00378, 0.00355]    & 2.0301$\times 10^{-17}$  \\
700   & [-0.00475, 0.00451]   & 2.5686$\times 10^{-17}$  & [-0.00349, 0.00326]    & 1.8712$\times 10^{-17}$  \\
1000  & [-0.00437, 0.00414]   & 2.3597$\times 10^{-17}$  & [-0.00321, 0.00298]    & 1.7156$\times 10^{-17}$ \\
\hline
 \cline{1-5}
\multicolumn{5}{|c|}{ $\sqrt{s}= 10$ TeV,  \hspace{1cm}  $95\%$ C.L.}\\
\hline
100   & [-0.00598, 0.00580]   & 3.2942$\times 10^{-17}$  & [-0.00450, 0.00431]    & 2.4691$\times 10^{-17}$  \\
300   & [-0.00472, 0.00454]   & 2.5946$\times 10^{-17}$  & [-0.00351, 0.00332]    & 1.9161$\times 10^{-17}$ \\
500   & [-0.00421, 0.00403]   & 2.3114$\times 10^{-17}$  & [-0.00312, 0.00293]    & 1.6979$\times 10^{-17}$  \\
700   & [-0.00391, 0.00372]   & 2.1391$\times 10^{-17}$  & [-0.00288, 0.00269]    & 1.5667$\times 10^{-17}$  \\
1000  & [-0.00360, 0.00341]   & 1.9685$\times 10^{-17}$  & [-0.00265, 0.00246]    & 1.4379$\times 10^{-17}$ \\
\hline\hline
\end{tabular}
\end{center}
\end{table}

\begin{table}[!ht]
\caption{Model-independent sensitivity estimate for the $\tilde{a}_\tau$ magnetic moment and the $\tilde{d}_\tau$ electric dipole
moment through the process $e^- e^+  \to e^+\,\tau^+ \tau^-\, \gamma e^-$ at CLIC.}
\begin{center}
\begin{tabular}{|c|cc|cc|}
\hline\hline
\multicolumn{5}{|c|}{ $\sqrt{s}= 1.5$ TeV,  \hspace{1cm}  $95\%$ C.L.}\\
 \hline
 \cline{1-5}
 & \multicolumn{2}{|c|}{Pure-leptonic} & \multicolumn{2}{|c|}{Semi-leptonic}\\
 \hline
 ${\cal L}\,(fb^{-1})$  & $\tilde{a}_\tau$ & \hspace{1.8cm}$ |\tilde{d}_\tau({\rm ecm})|$\hspace{1cm} & $\tilde{a}_\tau$  & \hspace{1.8cm}$ |\tilde{d}_\tau|({\rm ecm})$ \hspace{1cm} \\
\hline
100   & [-0.00505, 0.00470]  & 2.7222$\times 10^{-17}$  & [-0.00370, 0.00335]    & 1.9685$\times 10^{-17}$  \\
300   & [-0.00390, 0.00355]  & 2.0786$\times 10^{-17}$  & [-0.00286, 0.00252]    & 1.4996$\times 10^{-17}$ \\
500   & [-0.00346, 0.00311]  & 1.8322$\times 10^{-17}$  & [-0.00254, 0.00220]    & 1.3209$\times 10^{-17}$  \\
1000  & [-0.00294, 0.00259]  & 1.5430$\times 10^{-17}$  & [-0.00216, 0.00182]    & 1.1116$\times 10^{-17}$  \\
1500  & [-0.00267, 0.00233]  & 1.3953$\times 10^{-17}$  & [-0.00197, 0.00163]    & 1.0048$\times 10^{-17}$ \\
\hline
 \cline{1-5}
\multicolumn{5}{|c|}{ $\sqrt{s}= 3$ TeV,  \hspace{1cm}  $95\%$ C.L.}\\
\hline
100   & [-0.00385, 0.00362]   & 2.0650$\times 10^{-17}$  & [-0.00283, 0.00259]    & 1.4965$\times 10^{-17}$  \\
500   & [-0.00264, 0.00241]   & 1.5797$\times 10^{-17}$  & [-0.00194, 0.00171]    & 1.0055$\times 10^{-17}$ \\
1000  & [-0.00224, 0.00201]   & 1.1741$\times 10^{-17}$  & [-0.00165, 0.00142]    & 8.4650$\times 10^{-18}$  \\
2000  & [-0.00191, 0.00168]   & 9.8889$\times 10^{-18}$  & [-0.00141, 0.00118]    & 7.1239$\times 10^{-18}$  \\
3000  & [-0.00174, 0.00150]   & 8.9417$\times 10^{-18}$  & [-0.00128, 0.00105]    & 6.4394$\times 10^{-18}$ \\
\hline\hline
\end{tabular}
\end{center}
\end{table}


\vspace{1.5cm}

\begin{center}
{\bf Acknowledgments}
\end{center}

A. G. R. and M. A. H. R. acknowledge support from SNI and PROFOCIE (M\'exico).


\vspace{2cm}


\begin{figure}[H]
\centerline{\scalebox{0.8}{\includegraphics{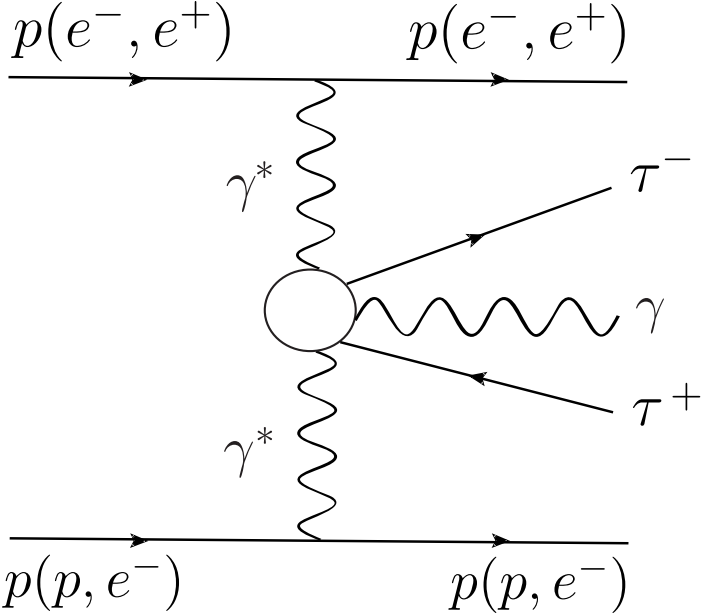}}}
\caption{ \label{fig:gamma1} A schematic diagram for the processes
$pp (e^-p, e^+e^-)  \to p\tau\bar \tau \gamma p (e^- \tau\bar \tau \gamma p, e^+ \tau\bar \tau \gamma e^-)$.}
\label{Fig.1}
\end{figure}

\begin{figure}[H]
\centerline{\scalebox{0.8}{\includegraphics{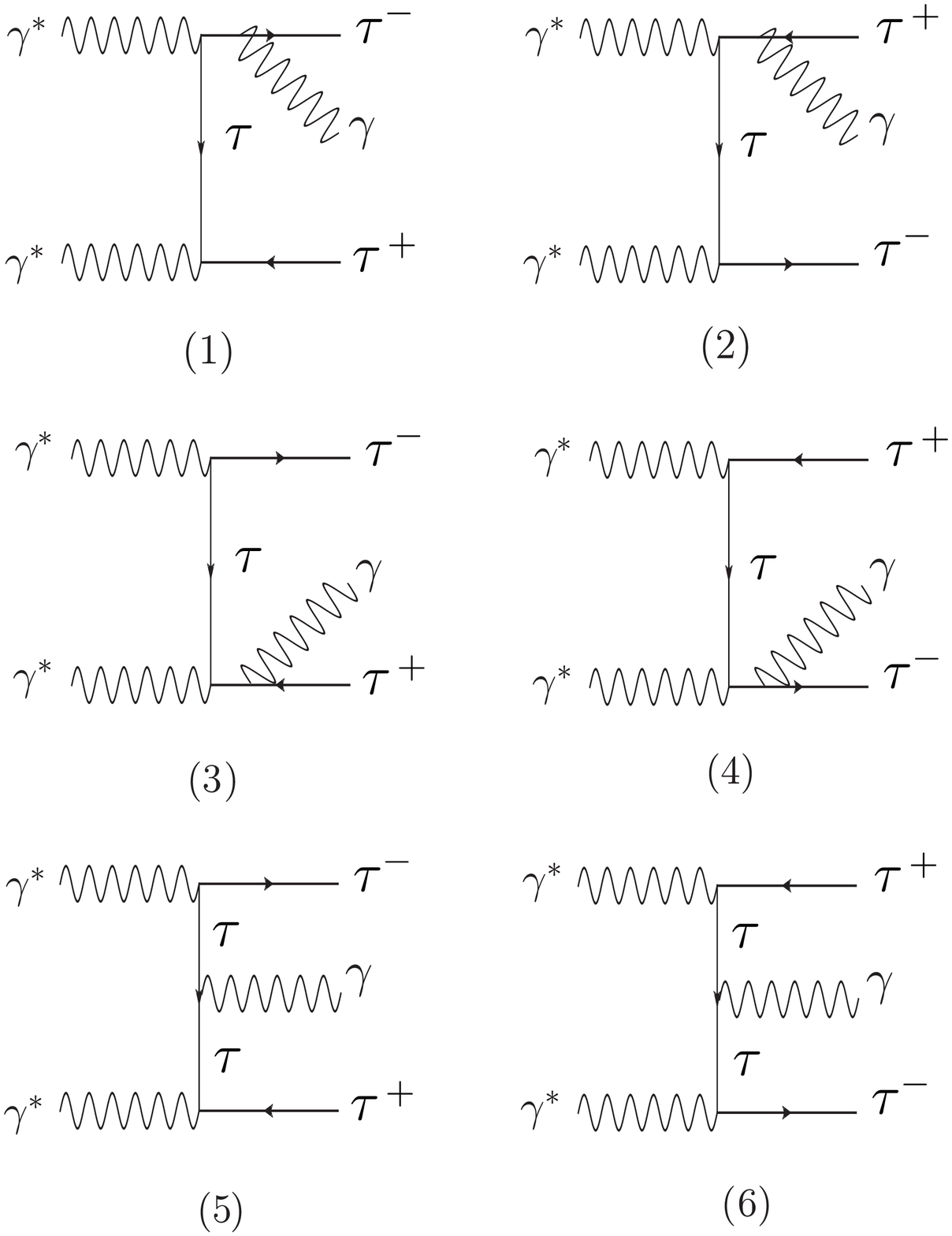}}}
\caption{ \label{fig:gamma2} Feynman diagrams contributing to the subprocess
$\gamma^*\gamma^* \to \tau \bar \tau \gamma$.}
\label{Fig.2}
\end{figure}

\begin{figure}[H]
\centerline{\scalebox{1.2}{\includegraphics{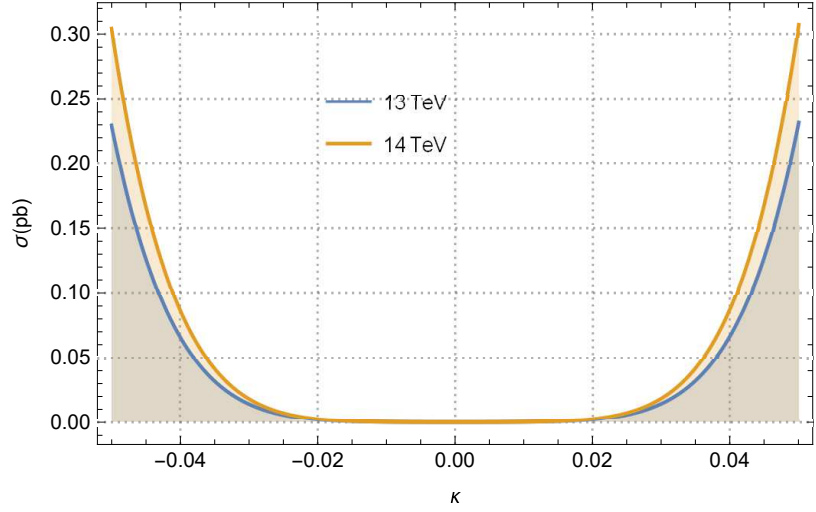}}}
\caption{The total cross-sections of the process
$pp \to p\tau\bar \tau \gamma p$ as a function of $\kappa$
for center-of-mass energies of $\sqrt{s}=13, 14$\hspace{0.8mm}$TeV$ at the LHC.}
\label{Fig.3}
\end{figure}

\begin{figure}[H]
\centerline{\scalebox{1.2}{\includegraphics{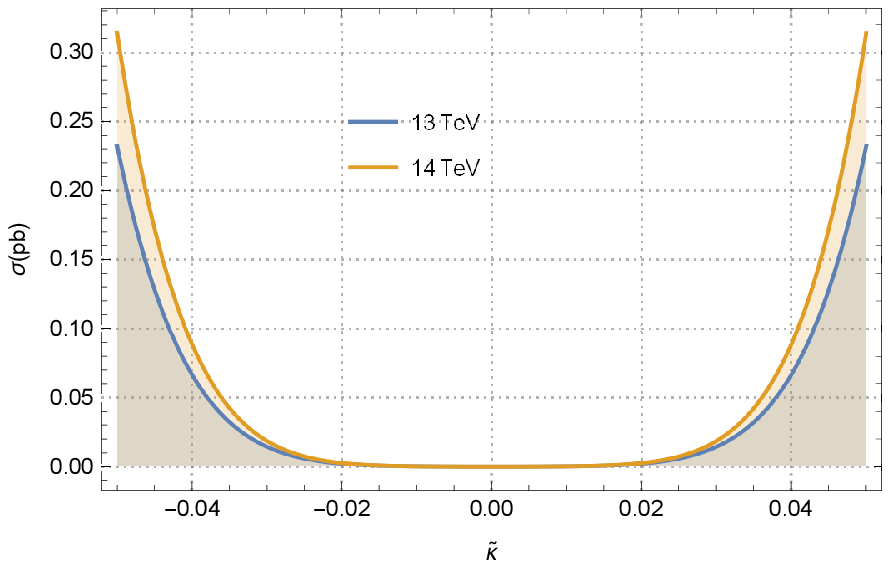}}}
\caption{ Same as in Fig. 3, but for $\tilde\kappa$.}
\label{Fig.4}
\end{figure}

\begin{figure}[H]
\centerline{\scalebox{1}{\includegraphics{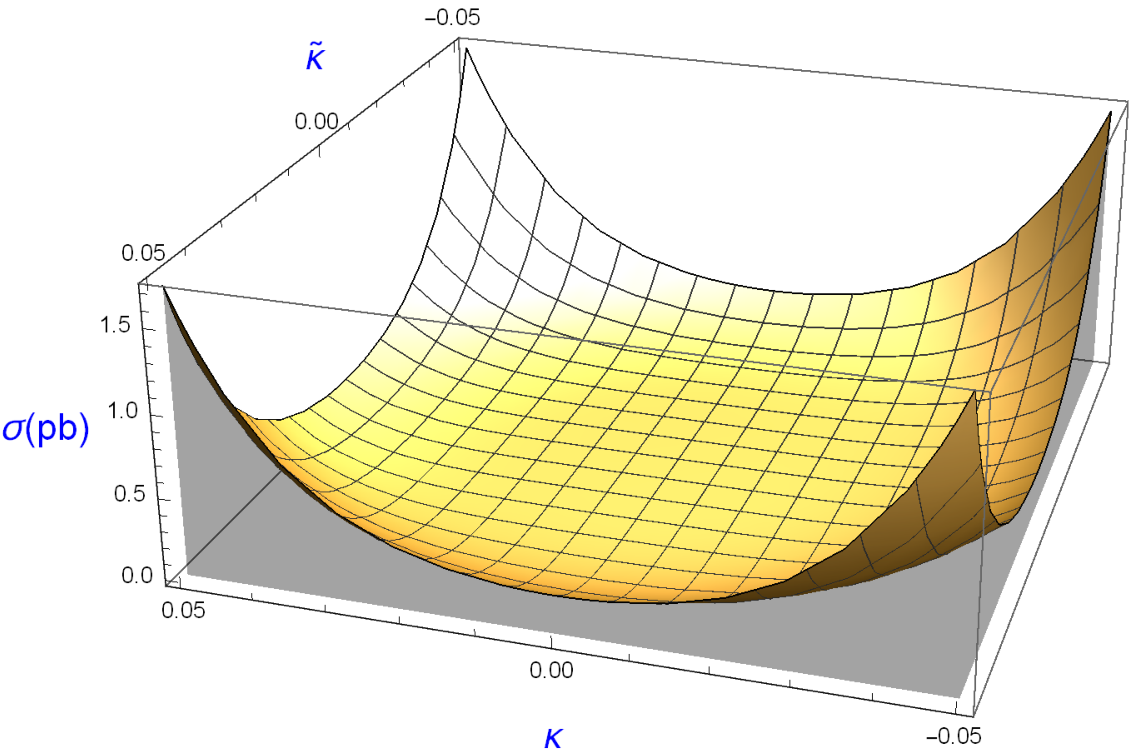}}}
\caption{ \label{fig:gamma1} The total cross-sections of the process
$pp \to p\tau\bar \tau \gamma p$ as a function of $\kappa$ and $\tilde\kappa$
for center-of-mass energy of $\sqrt{s}=13\hspace{0.8mm}TeV$ at the LHC.}
\label{Fig.5}
\end{figure}

\begin{figure}[H]
\centerline{\scalebox{1}{\includegraphics{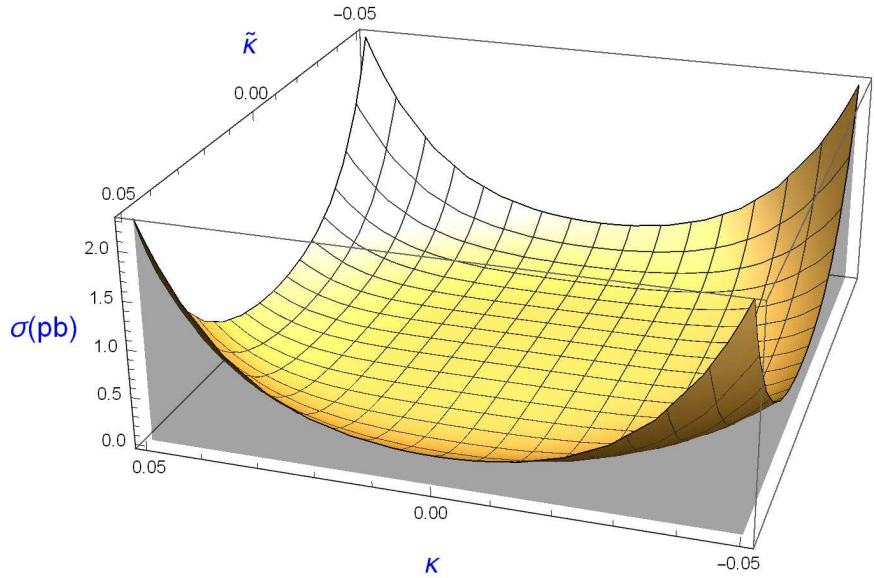}}}
\caption{ \label{fig:gamma2} Same as in Fig. 5, but for center-of-mass energy of
$\sqrt{s}=14\hspace{0.8mm}TeV$.}
\label{Fig.6}
\end{figure}

\begin{figure}[H]
\centerline{\scalebox{1}{\includegraphics{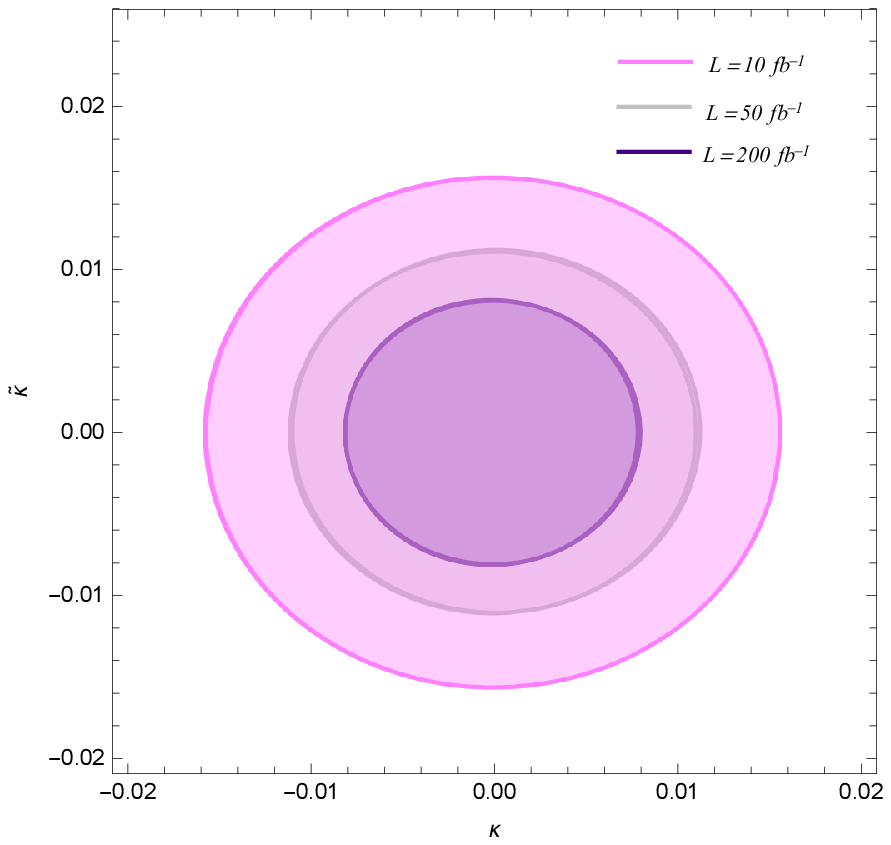}}}
\caption{ \label{fig:gamma1x} Sensitivity contours at the $95\% \hspace{1mm}C.L.$ in the
$\kappa-\tilde\kappa$ plane through the process $pp \to p\tau\bar \tau \gamma p$
for $\sqrt{s}=13$\hspace{0.8mm}$TeV$ at the LHC.}
\label{Fig.7}
\end{figure}

\begin{figure}[H]
\centerline{\scalebox{1}{\includegraphics{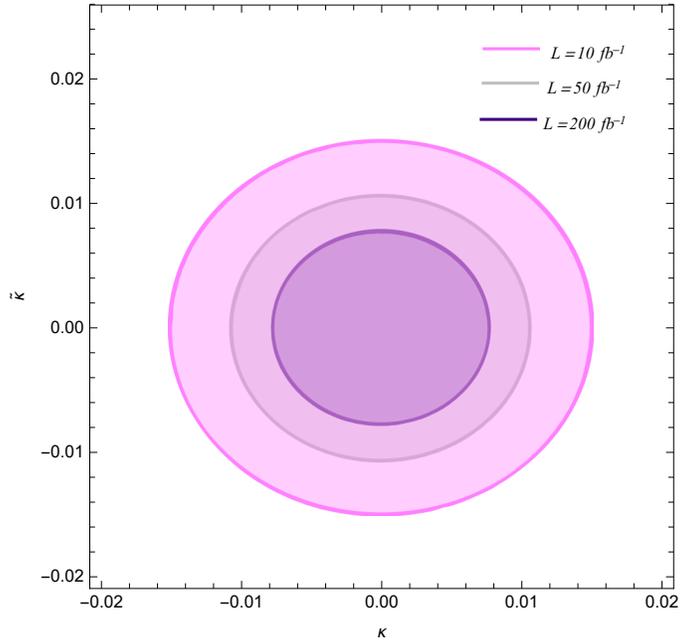}}}
\caption{ \label{fig:gamma2x} Same as in Fig. 7, but for $\sqrt{s}=14$\hspace{0.8mm}$TeV$.}
\label{Fig.8}
\end{figure}

\begin{figure}[H]
\centerline{\scalebox{1.2}{\includegraphics{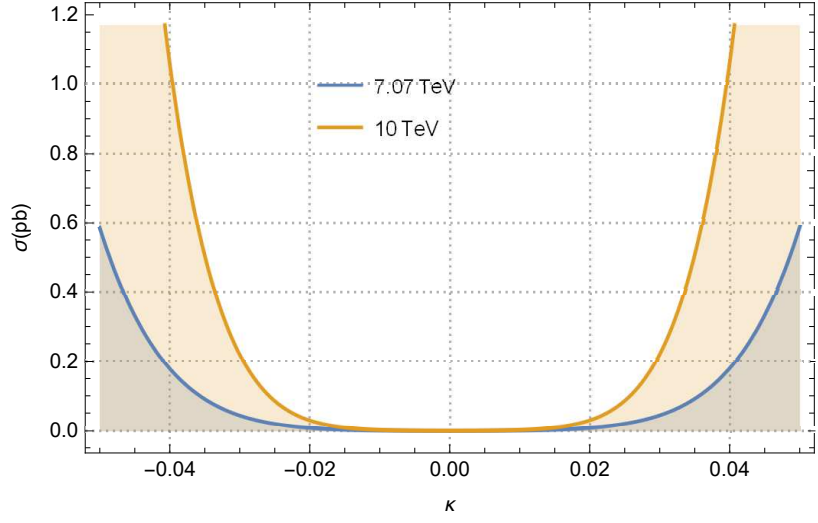}}}
\caption{ \label{fig:gamma15} The total cross-sections of the process
$e^-p \to e^- \tau\bar \tau \gamma p$ as a function of $\kappa$
for center-of-mass energies of $\sqrt{s}=7.07, 10$\hspace{0.8mm}$TeV$ at the FCC-he.}
\label{Fig.6}
\end{figure}

\begin{figure}[H]
\centerline{\scalebox{1.2}{\includegraphics{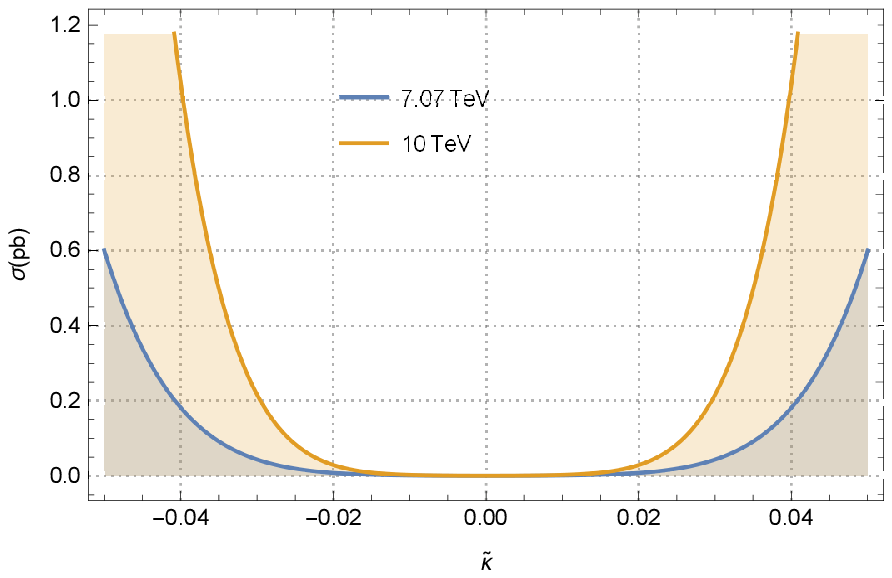}}}
\caption{ \label{fig:gamma6} Same as in Fig. 9, but for $\tilde\kappa$.}
\label{Fig.7}
\end{figure}

\begin{figure}[H]
\centerline{\scalebox{1}{\includegraphics{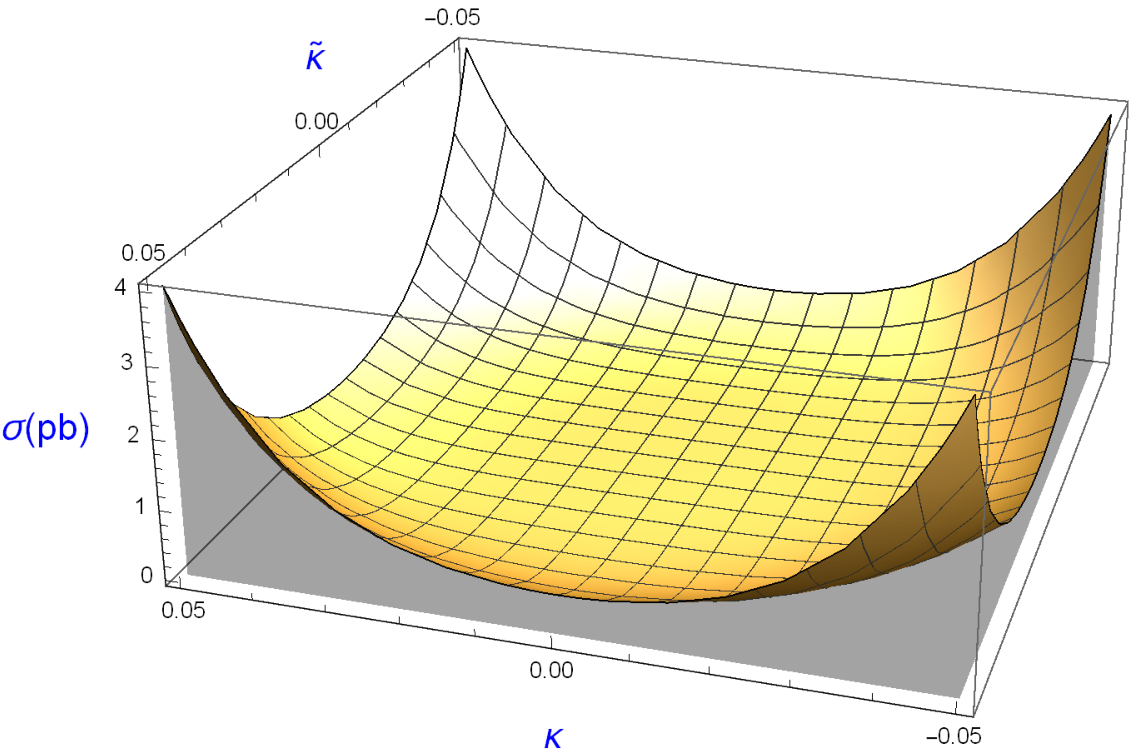}}}
\caption{ \label{fig:gamma6x} The total cross-sections of the process
$e^-p \to e^- \tau\bar \tau \gamma p$ as a function of $\kappa$ and $\tilde\kappa$
for center-of-mass energy of $\sqrt{s}=7.07\hspace{0.8mm}TeV$ at the FCC-he.}
\label{Fig.8}
\end{figure}

\begin{figure}[H]
\centerline{\scalebox{1}{\includegraphics{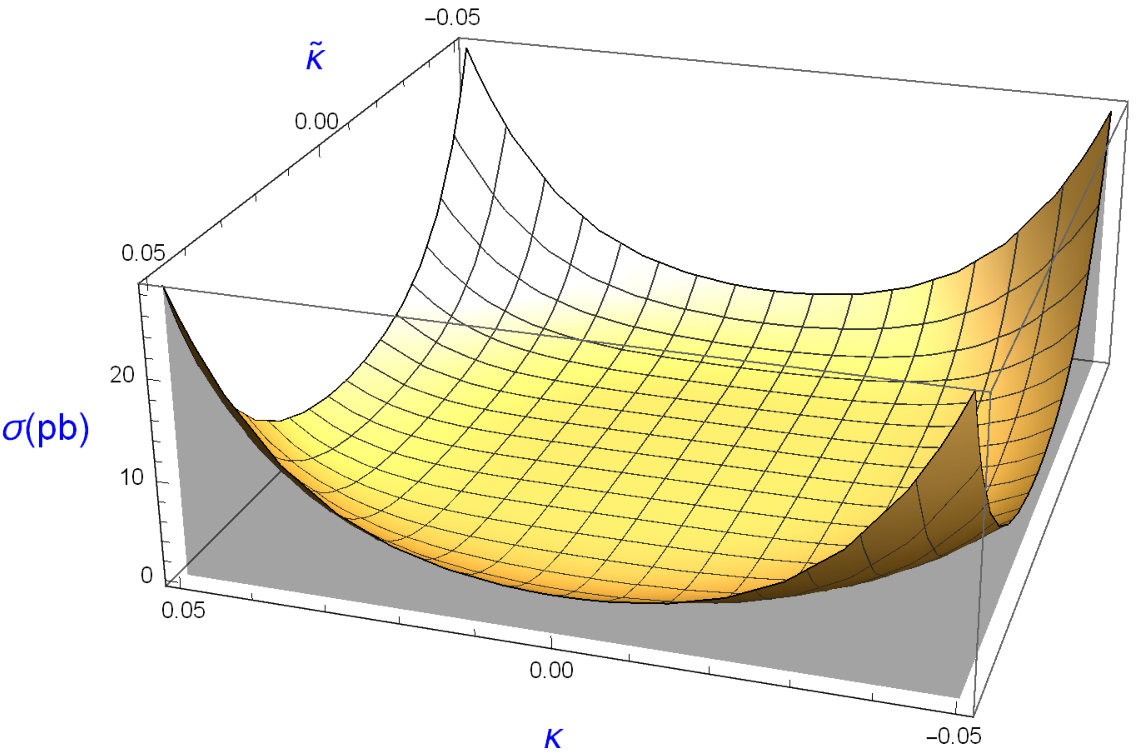}}}
\caption{ \label{fig:gamma6x} Same as in Fig. 11, but for
$\sqrt{s}=10\hspace{0.8mm}TeV$.}
\label{Fig.8}
\end{figure}

\begin{figure}[H]
\centerline{\scalebox{1.1}{\includegraphics{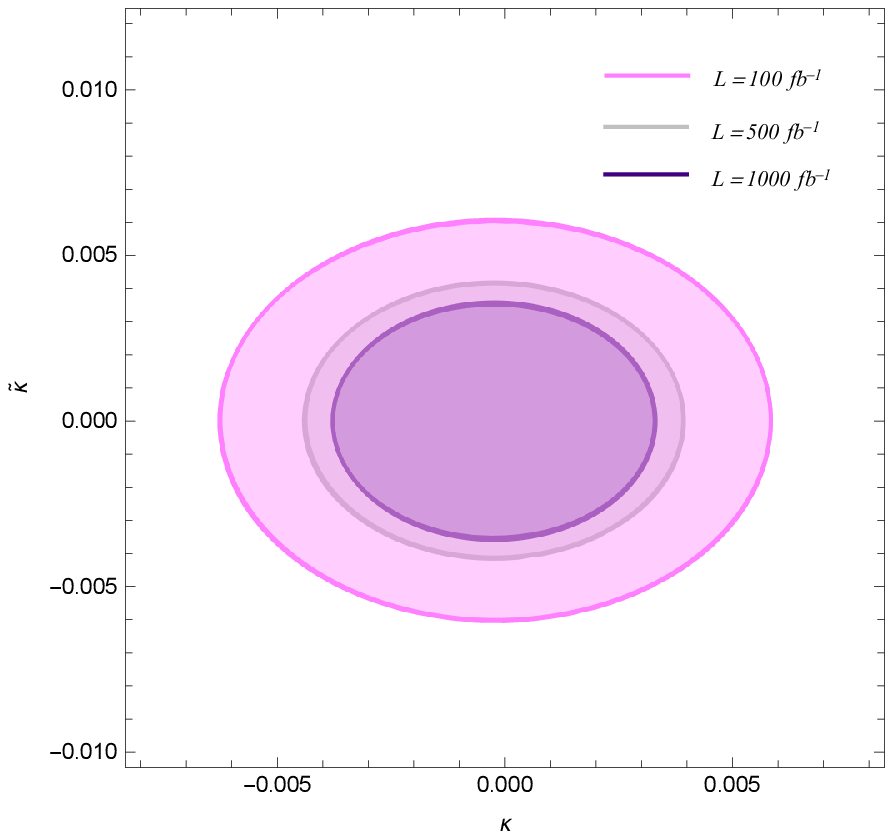}}}
\caption{ \label{fig:gamma1x} Sensitivity contours at the $95\% \hspace{1mm}C.L.$ in the
$\kappa-\tilde\kappa$ plane through the process $e^-p \to e^- \tau\bar \tau \gamma p$
for $\sqrt{s}=7.07$\hspace{0.8mm}$TeV$ at the FCC-he.}
\label{Fig.7}
\end{figure}

\begin{figure}[H]
\centerline{\scalebox{1.1}{\includegraphics{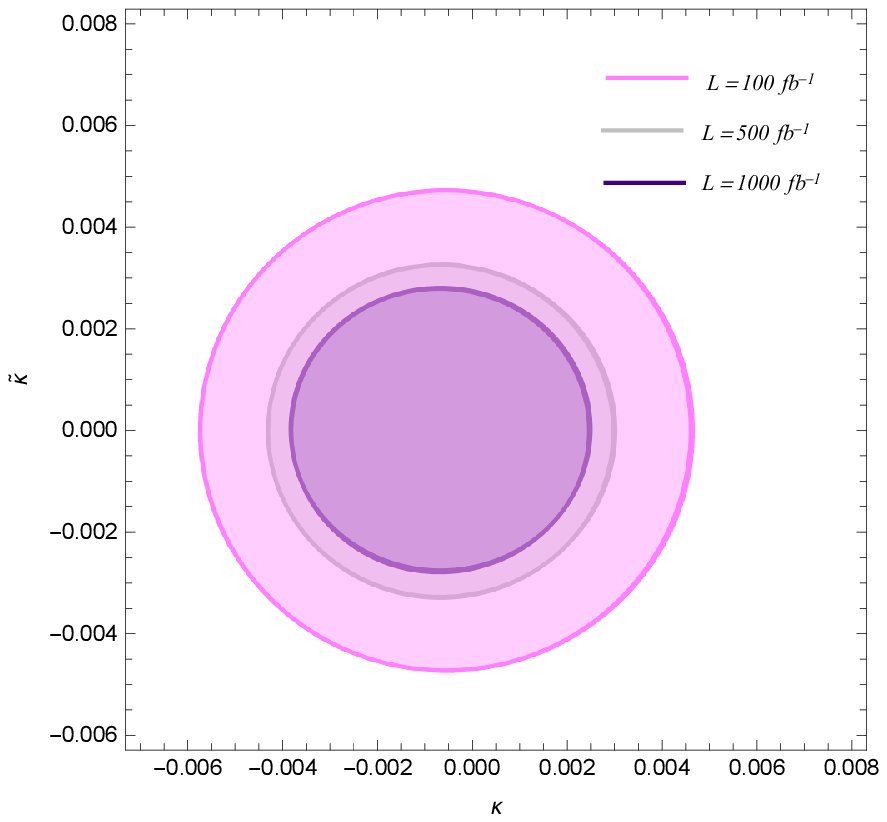}}}
\caption{ \label{fig:gamma2x} Same as in Fig. 13, but for $\sqrt{s}=10$\hspace{0.8mm}$TeV$.}
\label{Fig.8}
\end{figure}

\begin{figure}[H]
\centerline{\scalebox{1.2}{\includegraphics{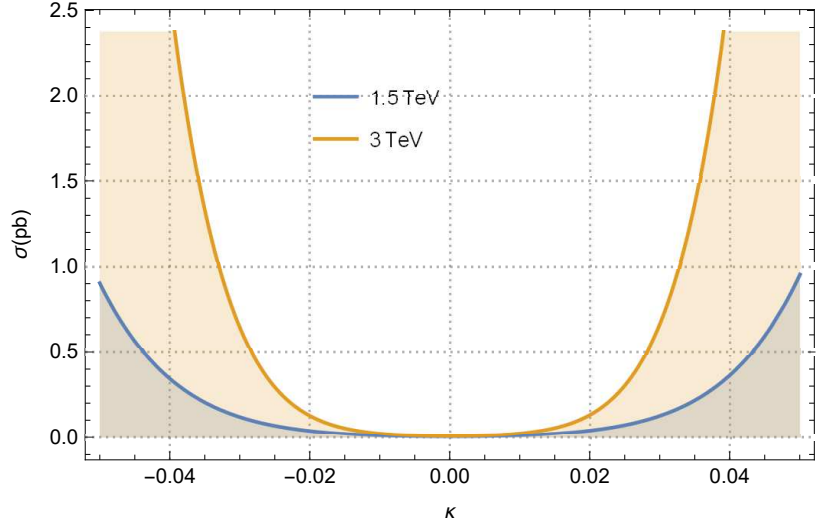}}}
\caption{ \label{fig:gamma15} The total cross-sections of the process
$e^+e^- \to e^+\tau\bar \tau \gamma e^-$ as a function of $\kappa$
for center-of-mass energies of $\sqrt{s}=1.5, 3$\hspace{0.8mm}$TeV$ at the CLIC.}
\label{Fig.6}
\end{figure}

\begin{figure}[H]
\centerline{\scalebox{1.2}{\includegraphics{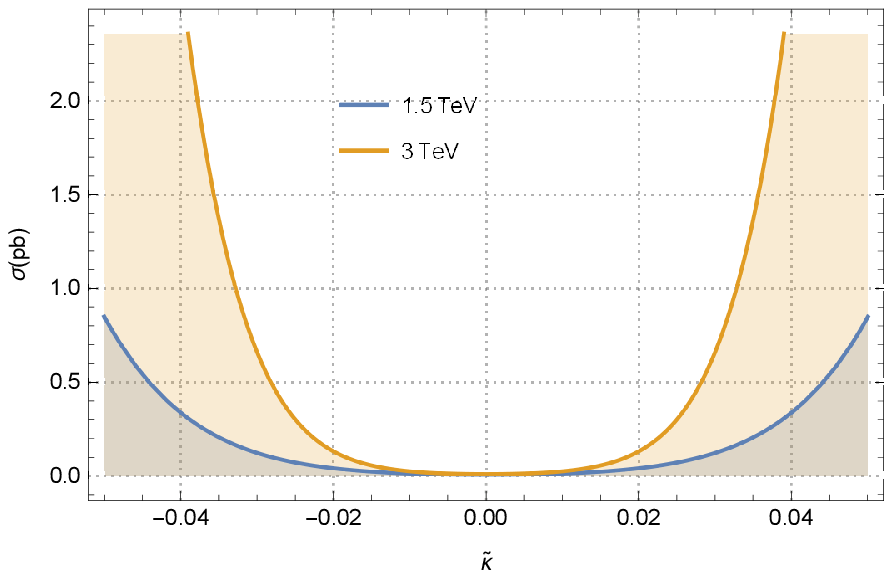}}}
\caption{ \label{fig:gamma6} Same as in Fig. 15, but for $\tilde\kappa$.}
\label{Fig.7}
\end{figure}

\begin{figure}[H]
\centerline{\scalebox{1}{\includegraphics{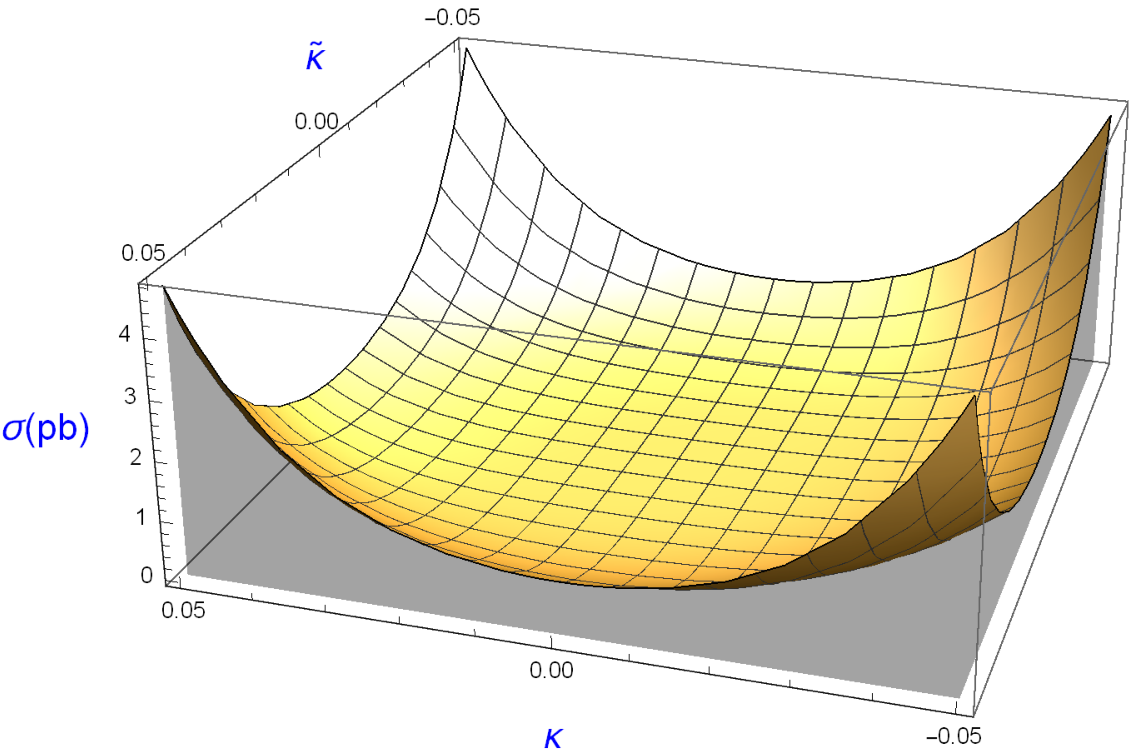}}}
\caption{ \label{fig:gamma6x} The total cross-sections of the process
$e^+e^- \to e^+\tau\bar \tau \gamma e^-$ as a function of $\kappa$ and $\tilde\kappa$
for center-of-mass energy of $\sqrt{s}=1.5\hspace{0.8mm}TeV$ at the CLIC.}
\label{Fig.8}
\end{figure}

\begin{figure}[H]
\centerline{\scalebox{1}{\includegraphics{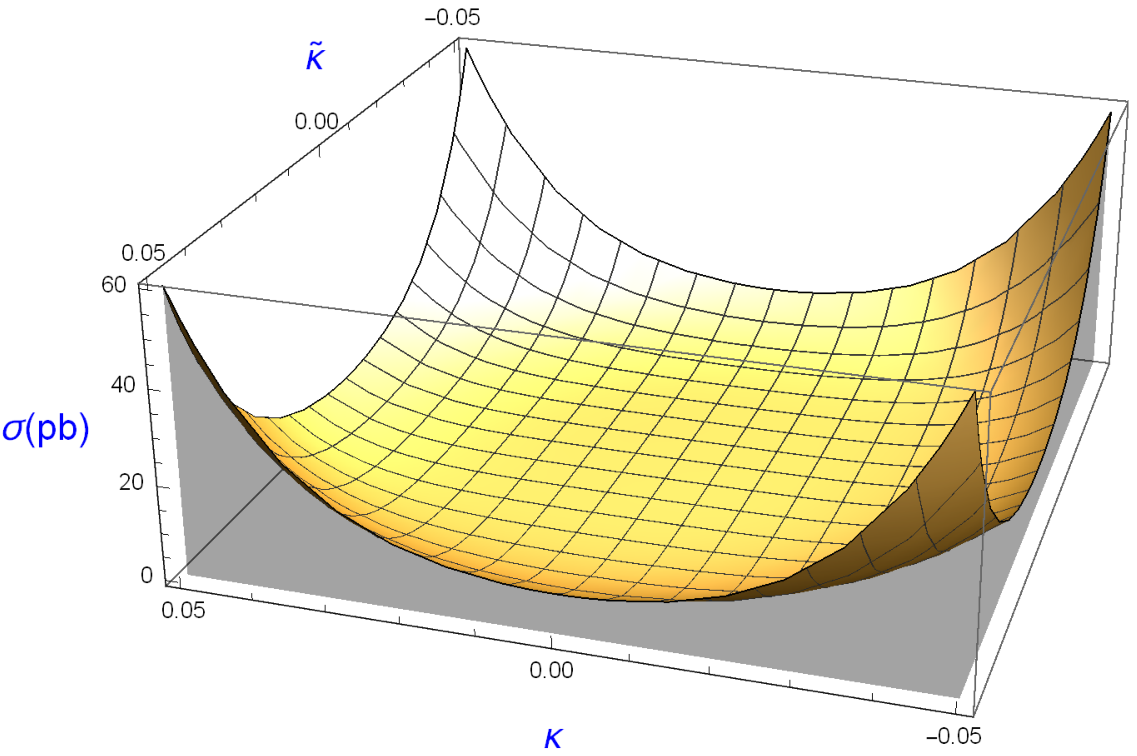}}}
\caption{ \label{fig:gamma6x} Same as in Fig. 17, but for $\sqrt{s}=3\hspace{0.8mm}TeV$.}
\label{Fig.8}
\end{figure}

\begin{figure}[H]
\centerline{\scalebox{1.1}{\includegraphics{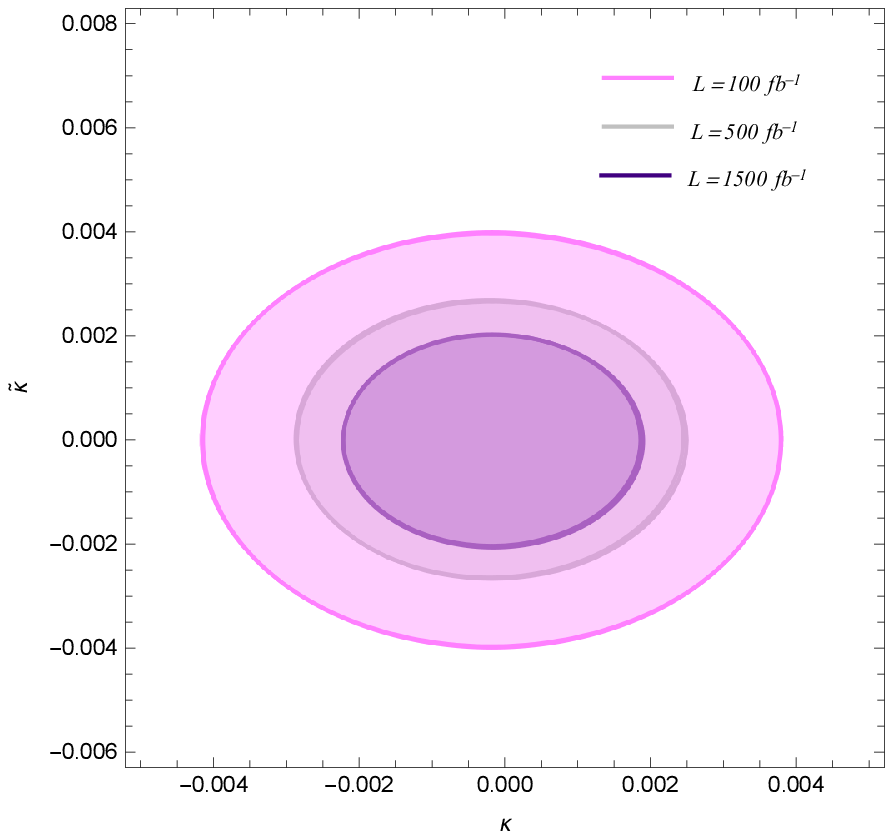}}}
\caption{ \label{fig:gamma1x} Sensitivity contours at the $95\% \hspace{1mm}C.L.$ in the
$\kappa-\tilde\kappa$ plane through the process $e^+e^- \to e^+\tau\bar \tau \gamma e^-$
for $\sqrt{s}=1.5$\hspace{0.8mm}$TeV$ at the CLIC.}
\label{Fig.7}
\end{figure}

\begin{figure}[H]
\centerline{\scalebox{1.1}{\includegraphics{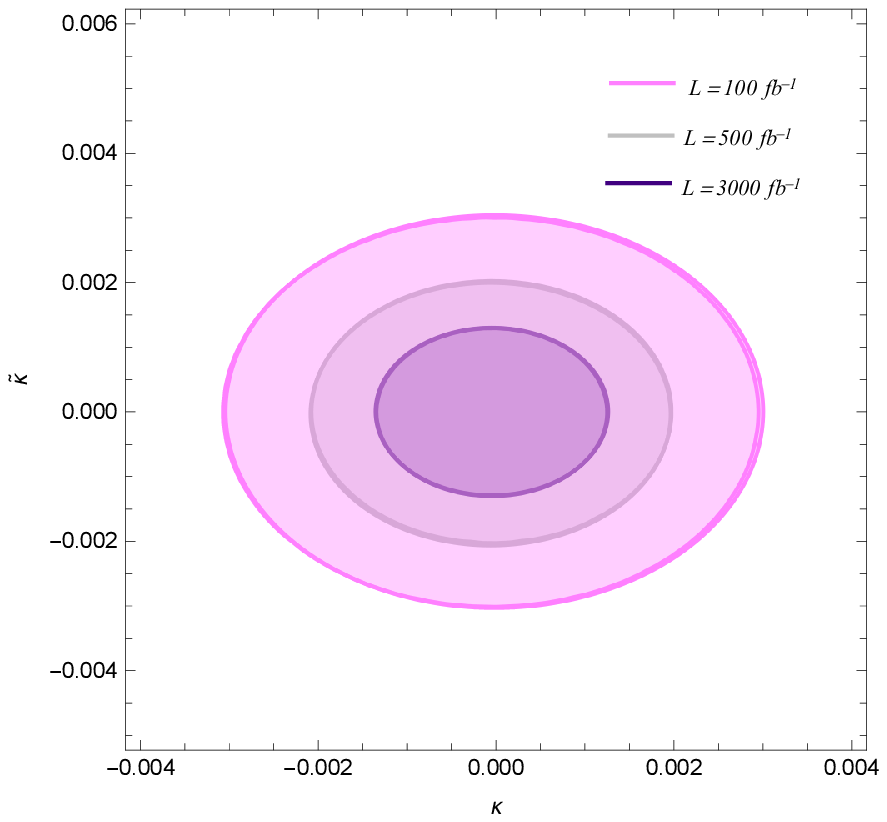}}}
\caption{ \label{fig:gamma2x} Same as in Fig. 19, but for $\sqrt{s}=3$\hspace{0.8mm}$TeV$.}
\label{Fig.8}
\end{figure}

\end{document}